\documentclass[MSNbibl,nameyear,seceqn,dvips]{arxstspdf}
\usepackage{dcolumn}
\usepackage{multirow}
\newcolumntype{d}[1]{D{.}{.}{#1}}
\usepackage{flushend}
\usepackage{stfloats}

\volume{27}
\issue{3}
\pubyear{2012}
\firstpage{412}
\lastpage{433}
\doi{10.1214/12-STS396} 

\makeatletter
\newproclaim{example}{Example}
\renewcommand{\epsilon}{\varepsilon}
\makeatother

\begin{document}
\begin{frontmatter}

\title{Models for Paired Comparison Data: A~Review with Emphasis on
Dependent~Data}
\runtitle{Models for Paired Comparison Data}

\begin{aug}
\author[a]{\fnms{Manuela} \snm{Cattelan}\corref{}\ead[label=e1]{manuela.cattelan@unipd.it}}
\runauthor{M. Cattelan}

\affiliation{Universit\`a di Padova}

\address[a]{Manuela Cattelan is Postdoctoral Research Fellow, Department of Statistical
Sciences, University of Padua,
via C. Battisti 241, 35121 Padova, Italy \printead{e1}.}

\end{aug}

%
\begin{abstract}
Thurstonian and Bradley--Terry models are the most commonly applied
models in the analysis of paired comparison data. Since their
introduction, numerous developments have been proposed in different
areas. This paper provides an updated overview of these extensions,
including how to account for object- and subject-specific covariates
and how to deal with ordinal paired comparison data. Special emphasis
is given to models for dependent comparisons. Although these models are
more realistic, their use is complicated by numerical difficulties. We
therefore concentrate on implementation issues. In particular, a
pairwise likelihood approach is explored for models for dependent
paired comparison data, and a simulation study is carried out to
compare the performance of maximum pairwise likelihood with other
limited information estimation methods. The methodology is illustrated
throughout using a real data set about university paired comparisons
performed by students.
\end{abstract}

%

\vspace*{-8pt}
\end{frontmatter}
%

\section{Introduction}\vspace*{1pt}

Paired comparison data originate from the comparison of objects in
couples. This type of data arises in numerous contexts, especially when
the judgment of a person is involved. Indeed, it is easier for people
to compare pairs of objects than ranking a list of items. There are
other situations that may be regarded as comparisons from which a
winner and a loser can be identified without the presence of a~judge.
Both these instances can be analyzed by the techniques described in
this paper.\vadjust{\goodbreak}

The objects involved in the paired comparisons can be beverages, carbon
typewriter ribbons, lotteries, players, moral values, physical stimuli
and many more. Here, the elements that are compared are called objects
or sometimes stimuli. The paired comparisons can be performed by a
person, an agent, a consumer, a~judge, et cetera, so the terms subject
or judge will be employed to denote the person that makes the choice.

The bibliography by \citet{Davidson76}, which includes more than 350
papers related to paired comparison data, testifies to the widespread
interest in this type of data. This interest is still present and
extensions of models for paired comparison data have been proposed.
This paper focuses on recent extensions of the two traditional models,
the \citet{Thurstone27} and the Bradley--Terry (\cite{Bradley52})
model, especially those subsequent to the review by \citet{Bradley76}
and the monograph by \citet{David88}, including in particular the work
that has been done in the statistical and the psychometric literature.\vadjust{\goodbreak}

Section \ref{sec2} reviews models for independent data. After the
introduction of the two classical models for the analysis of paired
comparison data and a~survey of different areas of application,
Sections \ref{sec3} and \ref{sec4} review extensions for ordinal paired
comparison data and for inclusion of explanatory variables. Section
\ref{sec5} reviews models that allow for dependence among the
observations and outlines the inferential problems related to such an
extension. Here, a pairwise likelihood approach is proposed to estimate
these models, and a simulation study is performed in order to compare
the estimates produced by maximum likelihood, a common type of limited
information estimation and pairwise likelihood. Section
\ref{sec6} reviews existing \texttt{R} (\cite{R}) packages for the
statistical analysis of paired comparison data, and Section~\ref{sec7} concludes.

\section{Independent Data}
\label{sec2}
\subsection{Traditional Models}

Let $Y_{sij}$ denote the random variable associated with the result of
the paired comparison between objects $i$ and $j$, $j>i=1, \ldots, n$,
made by subject $s=1, \ldots, S$, and let $\mathbf{Y}_{s}=(Y_{s12},
\ldots, Y_{s\,n-1\,n})$ be the vector of the results of all paired
comparisons made by subject $s$. When $S=1$ or the difference between
judges is not accounted for in the model, then the subscript $s$ will
be dropped. If each possible paired comparison is performed, they
number $N=n (n-1)/2$, and $SN=Sn (n-1)/2$ in a multiple judgment
sampling scheme, that is, when all paired comparisons are made by all
$S$ subjects. Different sampling schemes are possible. When each paired
comparison is performed by a different subject, the outcomes are
independent. In other instances, a subject performs more than one
paired comparison; in this case, it is conceivable that results of
several paired comparisons performed by the same subject will not be
independent. In Section \ref{sec2}, independence among observations is
assumed while Section \ref{sec5} addresses the issue of dependent data,
assuming that each subject performs all $N$ paired comparisons, except
for Section \ref{sec52} which considers the case of dependence not
induced by judges.

Let $\mu_{i} \in\mathbf{R}$, $i=1, \ldots,n$, denote the notional
worth of the objects. Traditional models were developed assuming only
two possible outcomes of each comparison, so $Y_{ij}$ is a binary
random variable, and $\pi_{ij}$, the probability that object $i$ is
preferred to object $j$, depends on the difference\vadjust{\goodbreak} between the worth of
the two objects
%
%
\begin{equation}\label{LinMod}
\pi_{ij}=F(\mu_{i}-\mu_{j}),
\end{equation}
where $F$ is the cumulative distribution function of a zero-symmetric
random variable. Such models are called linear models by
\citet{David88}. When $F$ is the normal cumulative distribution
function, formula (\ref{LinMod}) defines the \citet{Thurstone27}
model, while if $F$ is the logistic cumulative distribution function,
then the Bradley--Terry model (\cite{Bradley52}) is recovered. Other
specifications are possible; for example, \citet{Stern90} suggests
modeling the worth parameters as independent gamma variables with the
same shape parameter and different scale parameter. The Thurstone model
is also known as the Thurstone--Mosteller model since
\citet{Mosteller51} presented some inferential techniques for the
model, while the Bradley--Terry model was independently proposed also
by \citet{Zermelo29} and \citet{Ford57}. Model (\ref{LinMod}) is
called unstructured model, and the aim of the analysis is to make
inference on the vector $\bolds{\mu}=(\mu_{1}, \ldots,
\mu_{n})^{\prime}$ of worth parameters which can be used to determine a
final ranking of all the objects compared. Note that the specification
of model (\ref{LinMod}), through all the pairwise differences
$\mu_{i}-\mu_{j}$, implies that a constraint is needed in order to
identify the parameters. Various constraints can be specified: the most
common are the sum constraint,\break $\sum_{i=1}^{n} \mu_{i}=0$, and the
reference object constraint, $\mu_{i}=0$ for one object $i \in\{1,
\ldots,n\}$.

The comparative nature of the data poses inferential and
interpretational problems. Consider two different studies, for example,
about beverages. If subjects were requested to express an absolute
measure of like/dislike for each drink in a categorical scale, then the
data obtained from the two studies might be analyzed all together. On
the contrary, if the subjects express preferences in paired
comparisons, the data can be combined only if at least one object is
common to both studies; otherwise the data can be analyzed separately,
and no conclusions can be made about relationships between objects in
the two different studies. Indeed, the lack of origin implies that no
absolute statement can be made about the data and two subjects can
provide the same sets of preferences, but one may dislike all items
while the other may like all of them. The identification of an origin
may be useful for understanding the underlying psychological process,
for discriminating between desirable and undesirable objects and for
identifying the degree of an option desirability in different
conditions. However, it is not possible to recover the origin without
further choice experiments and/or further assumptions
(\cite{Thurstone57}; \cite{Bockenholt04}). Despite all their limits, paired
comparison data are widespread because of their ease of performance and
their discriminatory ability since objects that may be judged in the
same like/dislike category may be differentiated when compared
pairwise.

If the reference object constraint is employed, the identified worth
parameters are differences with respect to the reference object. Hence,
inference will typically regard differences between estimated worth
parameters with the related statistical problems. For example, for
testing $H_{0}\dvtx \mu_{i}=\mu_{j}$ by means of the Wald test statistic
$(\hat\mu_{i}- \hat\mu_{j})/\{\widehat{\operatorname{var}}(\hat\mu_{i}-
\hat
\mu_{j})\}^{1/2}$, where $\hat\mu_{i}$ is the maximum likelihood
estimator of $\mu_{i}$, the covariance between the estimators of the
worth parameters is needed. In general, the whole covariance matrix of
the worth parameters should be reported in order to allow the final
users to perform the tests they are interested in. However, it is very
inconvenient to report that matrix and a useful alternative may be to
report quasi-standard errors (\cite{Firth04}) instead of the usual
standard errors since they allow approximate inference on any of the
contrasts. Let $\mathbf{c}$ be a vector of zero-sum constants. If
the parameters $\bolds{\mu}$ were independent, then the estimated
standard error of $\mathbf{c}^{\prime} \bolds{\mu}$ would be $ (
\sum_{i=1}^{n} c_{i}^{2} \hat v_{i} ) ^{1/2}$, where $\hat v_{i}$
denotes the estimated variance of $\hat\mu_{i}$. Quasi-variances
are a vector of constants~$\mathbf{q}$ such that
\[
\operatorname{var} (\mathbf{c}^{\prime} \bolds{\mu} )\simeq
\sum_{i=1}^{n} c_{i}^{2} q_{i},
\]
so they have the property that they add over the components of
$\bolds{\mu}$, and hence can be used to approximate variances of
contrasts of estimated worth parameters as if they were independent.
Let $p(q_{i}+q_{j}, \widehat{\operatorname{var}}(\hat\mu_{i}- \hat\mu_{j}))$,
be a penalty function which depends on the quasi-variances and the
estimated variance of the difference $\hat\mu_{i}- \hat\mu_{j}$, then
quasi-variances are computed through minimization of the sum of the
penalty function for all contrasts; see Firth and de~Menezes (\citeyear{Firth04}, Section~2.1).

Further statistical problems arising from the comparative nature of the
data are discussed in Section~\ref{sec522}.

%
\begin{table}[b]
\caption{Universities paired comparison data. \texttt{1} and \texttt{2}
refer to~the~number of choices in favor of the university in~the~fist~and
the second column, respectively, while \texttt{X} denotes the
number of no preferences expressed}\label{tabuni}
\begin{tabular*}{\columnwidth}{@{\extracolsep{\fill}}llccd{3.0}@{\hspace*{-2pt}}}
\hline
& & $\mathbf{1}$ & $\bolds{X}$ & \multicolumn{1}{c@{}}{$\mathbf{2}$}\\
\hline
London & Paris & 186 & 26 & 91\\
London & Milan & 221 & 26 & 56\\
Paris & Milan & 121 & 32 & 59\\
London & St. Gallen & 208 & 22 & 73\\
Paris & St. Gallen & 165 & 19 & 119\\
Milan & St. Gallen & 135 & 28 & 140\\
London & Barcelona & 217 & 19 & 67\\
Paris & Barcelona & 157 & 37 & 109\\
Milan & Barcelona & 104 & 67 & 132\\
St. Gallen & Barcelona & 144 & 25 & 134\\
London & Stockholm & 250 & 19 & 34\\
Paris & Stockholm & 203 & 30 & 70\\
Milan & Stockholm & 157 & 46 & 100\\
St. Gallen & Stockholm & 155 & 50 & 98\\
Barcelona & Stockholm & 172 & 41 & 90\\
\hline
\end{tabular*}
\end{table}

\begin{example*}
A program supported by the European Union offers an
international degree in Economics and Management. Twelve universities
take part in this program, and in order to receive a degree, a student
in the program must spend a semester at another university taking part
in the program. Usually, some universities receive more preferences
than others, and this may cause organizational problems. A study was
carried out among 303 students of the Vienna University of Economics
who were asked in which university they would prefer to spend the
period abroad, between six universities situated in Barcelona (Escuela
Superior de Admi\-nistracion y~Direccion de Empresas), London (London
School of Economics and Political Sciences), Milan (Universit\`a Luigi
Bocconi), Paris (Hautes \'Etudes Commerciales), St. Gallen (Hochschule
St. Gallen) and Stockholm (Stockholm School of Economics), compared
pairwise. This example will be used throughout the paper as an
illustration. For an exhaustive analysis of the data refer to
Dittrich, Hatzinger and Katzenbeisser (\citeyear{Dittrich98}, \citeyear{Dittrich01}). The data set is available in both the
\texttt{prefmod} (\cite{Hatzinger10}) and the \texttt{BradleyTerry2}
(\cite{Turner10}) \texttt{R}~packages; see Section~\ref{sec6}. Table~\ref{tabuni} reports the aggregated data on the 15 paired comparisons.
For example, the first row shows that in the paired comparison between
London and Paris, 186 students prefer London, 91 students prefer Paris
and 26 students do not have a preference between the two universities.
Moreover, 91 students unintentionally overlooked the comparison between
Paris and Milan which has only 212 answers.\vadjust{\goodbreak} The second column of Table
\ref{tab1} shows the estimate of the worth parameters for the six
universities using the Thurstone model and adding half of the number of
no preferences to each university in the paired comparison. In Section~\ref{sec3}
a~better way to handle no preference data will be discussed.\looseness=1

The reference object constraint is used, and the worth parameter of
Stockholm is set to zero. All estimates are positive, so we can
conclude that Stockholm is the least preferred university, while London
is the most preferred one, followed by Paris,~Barce\-lona, St. Gallen
and Milan. The estimated probability that London is preferred to Paris
is $\Phi(0.982-0.561)=0.66$, where $\Phi$ denotes the cumulative
distribution function of a standard normal random variable. If it is of
interest to test whether the worth of St. Gallen is significantly
higher than the worth of Milan, the standard error of the difference
between these two worth parameters can be approximated by means of the
quasi-standard errors as $(0.030^{2}+0.031^{2})^{1/2}=0.043$.
Quasi-standard errors are lower than standard errors, thus accounting
for the positive covariance between parameter estimates. The value of
the test statistic is $(0.325-0.240)/0.043=1.98$, which yields a
$p$-value of 0.02; hence the hypothesis of equal worth parameters
between St. Gal\-len and Milan is not supported by the
data.
\end{example*}

%
\begin{table}
\tabcolsep=0pt
\caption{Estimates (\texttt{Est.}), standard errors (\texttt{S.E.}) and
quasi-standard errors (\texttt{Q.S.E.}) of the universities worth
parameters employing a two-categorical Thurstone model
(\texttt{Thurstone}) and a cumulative extension of the Thurstone model
(\texttt{cumulative Thurstone})}\label{tab1}
\begin{tabular*}{\columnwidth}{@{\extracolsep{4in minus 4in}}lcccccc@{}}
\hline
& \multicolumn{3}{c}{\textbf{Thurstone}} & \multicolumn{3}{c@{}}{\textbf{cumulative
Thurstone}}\\
 \ccline{2-4,5-7}\\ [-7pt]
& \textbf{Est.} & \textbf{S.E.} & \textbf{Q.S.E.} & \textbf{Est.} & \textbf{S.E.} & \textbf{Q.S.E.}\\
\hline
Barcelona & 0.333 & 0.043 & 0.030 & 0.332 & 0.041 & 0.028\\
London & 0.982 & 0.045 & 0.033 & 0.998 & 0.043 & 0.031\\
Milan & 0.240 & 0.044 & 0.031 & 0.241 & 0.041 & 0.029\\
Paris & 0.561 & 0.044 & 0.031 & 0.566 & 0.042 & 0.030\\
St. Gallen & 0.325 & 0.043 & 0.030 & 0.324 & 0.040 & 0.028\\
Stockholm & 0\phantom{.000} & -- & 0.031 & 0\phantom{.000} & -- & 0.029\\
$\tau_{2}$ & -- & -- & -- & 0.153 & 0.007 & --\\
\hline
\end{tabular*}
\vspace*{3pt}
\end{table}

\subsection{Applications}
There are many different areas in which paired comparison data arise.
Here, a number of recent applications are described, and further
references can be found in \citet{Bradley76}, \citet{Davidson76} and
\citet{David88}.

Despite its simplicity, the basic Bradley--Terry and Thurstone
models
have found a wide range of applications. \citet{Choisel07} analyze
pairwise evaluations of sounds through a standard Bradley--Terry model,
while \citet{Bauml94} and Kis\-sler and B\"{a}uml (\citeyear{Kissler00}) present applications
involving facial attractiveness. In \citet{Mazzucchi08} the standard
Bradley--Terry model is applied to a reliability problem. A panel of
wiring experts is asked to state which is the riskier one between
different scenarios compared pairwise in order to determine the
probability of wire failure as a~function of influencing factors in an
aircraft environment. \citet{Stigler94} uses the traditional
Bradley--Terry model for ranking scientific journals, and the same
model is exploited in genetics by \citet{Sham95}.

\citet{Maydeu08} list 10 reasons to use Thurstone's model for
analyzing subjective health outcomes, including the ease for
respondents, the existence of extensions for modeling inconsistent
choices and for including covariates and the possibility to investigate
which aspects influence the choices of subjects.

In many applications there are more than two possible outcomes of the
comparisons. \citet{Henery92} employs a Thurstone model for ranking
chess players and adapts it to three possible results: win, draw and
loss. \citet{Bockenholt97} consider a five-response-categories model
for applications to taste testing of beverages and to preferences for
brands of cigarettes. \citet{Dittrich04} consider motives to start a
Ph.D. program using three response categories in the log-linear version
of the Bradley--Terry model.

It is often of interest to investigate whether some covariates affect
the results of the comparisons. \citet{Ellermeier04} employ
a
Bradley--Terry model to analyze pairwise evaluations of sounds and
include sound-related covariates, for example, roughness, sharpness, et
cetera, to evaluate which of them contribute to the unpleasantness of
sounds. \citet{Duineveld00} use the log-linear formulation of the
Bradley--Terry model to investigate consumer preference data on orange
soft drinks including an analysis of the factorial design for the
drinks compared, while \citet{Francis02} include subject-specific
covariates in the analysis of value orientation of people in different
European countries. Applications of the Bradley--Terry model are
present also in zoological data in order to investigate\vadjust{\goodbreak} aspects of
animal behavior considering animal-specific covariates
(\cite{Stuart06}; \cite{Whiting06}; \cite{Head08}). \citeauthor{Agresti02} [(\citeyear{Agresti02}), Chapter 10]
extends the Bradley--Terry model to account for the home advantage
effect in baseball data.

Sometimes it is more realistic to include dependence among
observations. Object-specific random effects can be used to introduce
correlation between comparisons with common objects, for example, in
sports data (\cite{Cattelan09}). When all judges perform all paired
comparisons, random effects can introduce correlation between
preferences expressed by the same subject involving a common object as
shown in \citet{Bockenholt07} for the university preference data.

When paired comparisons are performed in prolonged time periods, it may
be necessary to account for it. \citet{McHale11} estimate a
Bradley--Terry model in which tennis matches distant in time are
down-weighted since the aim is to predict the results of future
matches. Further dynamic extensions for sports data have been proposed
by \citet{Barry93}, \citet{Fahrmeir94}, \citet{Held00} and
\citet{Cattelan10}. In tournaments it may happen that a player wins
all the comparisons in which he is involved. In this case a standard
Bradley--Terry or Thurstone model would estimate an infinity worth
parameter for this team. \citet{Mease03} proposes a penalization of
the likelihood which overcomes this problem. The meth\-od proposed by
\citet{Firth93} to reduce the bias of the maximum likelihood estimates
is an alternative technique to obtain finite estimates in this
instance. Finally, the case in which the margin of victory in sport
contests is not discrete, but continuous, is analyzed in
\citet{Stern11}.

In the context of the log-linear specification of the Bradley--Terry
model, \citet{Dittrich11} account also for missing responses in a
study about the qualities of a good teacher.

\subsection{Ordinal Paired Comparisons}\label{sec3}

Sometimes subjects are requested to express a degree of preference.
Suppose that objects $i$ and $j$ are compared, and the subject can
express strong preference for $i$ over $j$, mild preference for $i$, no
preference, mild preference for~$j$ over $i$ or strong preference for~$j$.
If $H$ denotes the number of grades of the scale, then in this
example, $H=5$.

Let $Y_{ij}=1, \ldots, H$, where 1 denotes the least favorable response
for $i$, and $H$ is the most favorable response for $i$.\vadjust{\goodbreak}
\citet{Agresti92} shows how two models for the analysis of ordinal
data can be adapted to ordinal paired comparison data. The {\em
cumulative link} models exploit the latent random variable
representation. Let $Z_{ij}$ be a continuous latent random variable,
and let $\tau_{1} < \tau_{2}< \cdots< \tau_{H-1}$ denote thresholds
such that $Y_{ij}=h$ when $\tau_{h-1}< Z_{ij} \leq\tau_{h}$. Then,
%
%
\begin{equation}
\label{cumulativeModel} \operatorname{pr}(Y_{ij} \leq
y_{ij})=F(\tau_{y_{ij}}-\mu_{i}+\mu_{j}),
\end{equation}
where $-\infty=\tau_{0} < \tau_{1} < \cdots<
\tau_{H-1}<\tau_{H}=\infty$, and $F$ is the cumulative distribution
function of the latent variable $Z_{ij}$. $F$ is usually assumed to be
either the logistic or the normal distribution function leading to the
cumulative logit or the cumulative probit model, respectively. The
symmetry of the model imposes that $\tau_{h} = -\tau_{H-h}$, $h=1,
\ldots, H$ and $\tau_{H/2}=0$ when~$H$ is even. When $H=3$ there are
two threshold parameters, $\tau_{1}$ and $\tau_{2}$, such that
$\tau_{1}=-\tau_{2}$ and model (\ref{cumulativeModel}) corresponds to
the extension of the Bradley--Terry model introduced by \citet{Rao67}
when a logit link is considered, and the extension of the Thurstone
model by \citet{Glenn60} when the probit link is employed.

An alternative model proposed by \citet{Agresti92} is the
\textit{adjacent categories} model. In this case the link is applied to
adjacent response probabilities, rather than cumulative probabilities
and reduces to the Bradley--Terry model when only 2 categories are
allowed and to the model proposed by \citet{Davidson70} when 3
categories are allowed. The adjacent categories model is simpler to
interpret than cumulative link models since the odds ratio refers to
a~given outcome instead of referring to groupings of outcomes
(\cite{Agresti92}). The adjacent categories model, as well as the
Bradley--Terry model, has also a log-linear representation
(\cite{Dittrich04}).

An application of the adjacent categories model to market data is
illustrated in \citet{BockenholtDillon97}.
\citet{Bockenholt97} note
that a bias may be caused by the usage of the scale because subjects
may use only subsets of all categories. The threshold parameters
$\tau_{h}$ can account for the selection bias, for example, in the
cumulative probit model the quantity $\Phi(\tau_{h})-\Phi(\tau_{h-1})$
gives the category selection bias since it is the probability of
selecting category $h$ when the two stimuli are equal. Different latent
classes of consumers with different threshold values and worth
parameters can be identified. If subjects share the same worth
parameters but have different thresholds, it is possible to let
thresholds depend on subject-specific covariates and to have a random
part (\cite{Bockenholt01b}). It is also possible to define thresholds
that depend on the objects compared, as in \citet{Henery92}.

\begin{example*}
In the paired comparisons of universities, students
were allowed to express no preference between two universities.
Therefore, the data should be analyzed by means of a model for ordinal
data. Columns 5--7 in Table \ref{tab1} show the estimates of a
cumulative probit extension of the Thurstone model for the university
data. The estimated threshold parameter $\hat\tau_{2}=0.153$ is highly
significant. In this particular case, the estimates of the worth
parameters and their standard errors are very similar to those of the
model with two categories, and the ranking of universities remains the
same, but in general, especially when the number of no preferences is
large, results can be different. Moreover, in this case it is possible
to estimate the probability of no preference between London and Paris
which is $\Phi(0.153-0.998+0.566)-\Phi(-0.153-0.998+0.566)=0.11$, and
the estimated probability that London is preferred to Paris reduces to
$1-\Phi(0.153-0.998+0.566)=0.61$; hence the estimated probability that
Paris is preferred to London is $0.28$. There is no much difference
from the previous result in the test of equality of worth parameters
for universities in St. Gallen and Milan.
\end{example*}

\subsection{Explanatory Variables}
\label{sec4}

In many instances, it is of interest to investigate whether some
explanatory variables affect the results of the comparisons.
Explanatory variables can be related to the objects compared, to the
subjects performing the comparisons or they can be compar\-ison-specific.

Let $\mathbf{x}_{i}=(x_{i1}, \dots, x_{iP})^{\prime}$ be a vector of
$P$ explanatory variables related to object $i$ and $\bolds{\beta
}=(\beta_{1}, \ldots, \beta_{P})$ be a $P$-dimensional parameter
vector. Then, in the context of the Bradley--Terry model,
\citet{Springall73} proposes to describe the worth parameters as the
linear combination
%
%
\begin{equation}\label{cov}
\hspace*{10pt}\mu_{i}= x_{i1} \beta_{1}+ \cdots+x_{iP} \beta_{P},
\quad i=1, \ldots, n.
\end{equation}
A paired comparison model with explanatory variables is called a
structured model. The same extension can be applied to the Thurstone
model. Note that since only the differences
$\mu_{i}-\mu_{j}=(\mathbf{x}_{i}- \mathbf{x}_{j})^{\prime}
\bolds{\beta}$ enter the linear predictor, an intercept cannot be
identified. In some instances, both worth parameters of objects and
further object-specific covariates are included, hence the linear
predictor assumes the form $\mu_{i}-\mu_{j}+(\mathbf{x}_{i}-
\mathbf{x}_{j})^{\prime} \bolds{\beta}$; see \citet{Stern11}.\vadjust{\goodbreak}

Model (\ref{cov}) has been extended to more flexible models, such as
additive combinations of spline\break smoothers (\cite{DeSoete93}); however
large data sets may be necessary to estimate nonlinearities reliably,
even though there is no investigation about this issue.

In case worth parameters are specified as in (\ref{cov}), standard
errors for the worth parameters can be computed through the delta
method, while when both the worth parameters and covariates are
included in the linear predictor, quasi-standard errors can be computed
for the worth parameters.

The results of the comparisons can be influenced also by
characteristics of the subject that performs the paired comparisons. In
the log-linear representation of the Bradley--Terry model,
\citet{Dittrich98} show how to include categorical subject specific
covariates, while \citet{Francis02} tackle the problem of continuous
subject-specific covariates and consider also the case in which some of
these covariates have a smooth nonlinear relationship.

\citet{Dillon93} consider a~marketing application and divide subjects
in latent classes to which they belong with a probability that depends
on their explanatory variables.

Covariates can be added in the linear predictor (\ref{cov}) if they are
subject-object interaction effects. For example, the knowledge of a
foreign language may influence the preference for a university. An
interaction effect can account for whether the student knows, for
example, Spanish and one object in the comparison is the university in
Barcelona. Unfortunately, subject covariates that do not interact with
objects, such as age of respondents, cannot be included.

A semiparametric approach which accounts for subject-specific
covariates is proposed by Strobl,\break Wickelmaier and Zeileis (\citeyear{Strobl10})
who suggest a meth\-odology
to partition recursively the subjects that perform the paired
comparisons on the basis of their covariates. The procedure tests
whether structural changes in the parameters occur for subjects with
different values of the covariates. Subjects are split according to the
test and a different unstructured Bradley--Terry model is fitted for
each subgroup. The method allows us to identify which covariates
influence the worth parameters without the need to assume a model for
them and finds the best cut point in case of continuous covariates.
Moreover, it is possible to include subject-specific covariates, not
only\vadjust{\goodbreak} interaction effects. Attention is needed in setting the minimum
number of subjects per class and in setting the significance level of
the test in order to avoid overfitting for large data sets. Differently
from the usual latent class models, the method allows to divide
subjects on the basis of their covariates; however, if some important
subject-specific covariates are not available, it may be expected that
the usual latent class model will perform better. In \citet{Strobl10}
an unstructured Bradley--Terry model is estimated for each subgroup,
but it seems possible to extend the method also to structured models.

Finally, there may be also comparison-specific covariates which are
related to the objects, but change from comparison to comparison. An
example of\break a~comparison-specific covariate is the home advantage effect
in sport tournaments since it depends on whether one of the players
competes in the home field. This effect may be accounted for by adding
a further term in the linear predictor (\ref{cov}). Another example is
the experience effect in contests between animals which, in
\citet{Stuart06}, is accounted for through a covariate that counts the
number of previous contests fought by animals.

\begin{example*}
In the universities paired compari\-sons, it may be of
interest to assess whether some object-specific covariates influence
the results of the comparisons. The universities in London and Milan
specialize in economics, the universities in Paris and Barcellona
specialize in management science and the remaining two in finance. This
aspect may influence the decisions of students. Another element that
may affect the comparisons is the location of the universities, in this
respect they can be divided in universities in Latin countries (Italy,
France and Spain) and universities in other countries.

Some features of the students that performed the universities paired
comparisons were collected, too. In particular, it is known whether
students have good knowledge of English, Italian, Spanish and\break French
and which is the main topic of their studies. It is conceivable that,
for example, students with a~good knowledge of French are more inclined
to prefer the university in Paris. Table \ref{tab4} shows the estimates
of a model with a linear predictor that includes object specific
covariates and subject-object interaction effects. Universities in
non-Latin countries are preferred to those in Latin countries, and
universities that specialize in finance seem less appealing to
students. The good knowledge of a foreign language induces students
to\vadjust{\goodbreak}
choose the university situated in the country where that foreign
language is spoken. Consider a student with a good knowledge of both
English and French and whose main discipline of study is management,
then the estimated probability that this student prefers London to
Paris is $1-\Phi\{0.160-(0.141+0.757-0.652-0.789+0.835-0.238)\}=0.46$,
while the estimated probabilities of no preference and preference for
Paris are 0.13 and 0.41, respectively. If this student's main
discipline of study was not management, which is the subject in which
Paris specializes, then the above estimated probabilities of preferring
London, no preference and preferring Paris would become 0.55, 0.12 and
0.33, respectively.
\end{example*}

%
\begin{table}
\caption{Estimates (\texttt{Est.}) and standard errors (\texttt{S.E.})
of universities data with subject- and object-specific covariates}\label{tab4}
\begin{tabular*}{\columnwidth}{@{\extracolsep{\fill}}ld{2.3}c@{}}
\hline
& \multicolumn{1}{c}{\textbf{Est.}} & \multicolumn{1}{c@{}}{\textbf{S.E.}}\\
\hline
Economics & 0.757 & 0.066\\
Management & 0.789 & 0.080\\
Latin country & -0.835 & 0.071\\
Discipline:Management & 0.238 & 0.054\\
English:London & 0.141 & 0.075\\
French:Paris & 0.652 & 0.049\\
Italian:Milan & 1.004 & 0.094\\
Spanish:Barcelona & 0.831 & 0.095\\
$\tau_{2}$ & 0.160 & 0.007\\
\hline
\end{tabular*}
\end{table}

\section{Models for Dependent Data}\label{sec5}
\subsection{Intransitive Preferences}

The models presented so far are estimated assuming independence among
all observations. The inclusion of a dependence structure is not only
more realistic, but also has an impact on the transitivity properties
of the model. Intransitive choices occur when object $i$ is preferred
to $j$, and object $j$ is preferred to $k$, but in the paired
comparison between $i$ and $k$, the latter is preferred. These are also
called circular triads. Paired comparison models can present different
transitivity properties. Assume that $\pi_{ij} \geq0.5$ and $\pi_{jk}
\geq0.5$, then a model satisfies:
\begin{itemize}
\item weak stochastic transitivity if $\pi_{ik} \geq0.5$;
\item
moderate stochastic transitivity if $\pi_{ik}
\geq\min(\pi_{ij},
\pi_{jk})$;
\item strong stochastic transitivity if $\pi_{ik} \geq
\max(\pi_{ij}, \pi_{jk})$.
\end{itemize}
The Bradley--Terry and Thurstone models as presented so far satisfy
strong stochastic transitivity. This property may be desirable
sometimes, for example, when asking wiring experts which\vadjust{\goodbreak} is the risk\-ier
situation between different scenarios in an aircraft environment. In
this case it is desirable that choices are consistent, so
Mazzucchi, Linzey and\break Bruning (\citeyear{Mazzucchi08}) use transitivity to check the level of reliability
of experts. However, in some situations choices can be systematically
intransitive, for example, when the same objects have more than one
aspect of interest, and different aspects prevail in different
comparisons.

\citet{Causeur05} propose a two-dimen\-sional Bradley--Terry model in
which the worth parameter of each object is bidimensional and can thus
be represented on a plane. A further multidimensional extension is
proposed by \citet{Usami10}. However, this methodology does not
provide a final ranking of all objects.

A different method that allows the inclusion in the model even of
systematic intransitive comparisons while yielding a ranking of all the
objects consists of modeling the dependence structure among
comparisons. The development of inferential techniques for dependent
data has recently allowed an investigation of models for dependent
observations.

\subsection{Multiple Judgment Sampling}
The assumption of independence is questioned in the case of the
multiple judgment sampling, that is, when $S$ people make all the $N$
paired comparisons. It seems more realistic to assume that the
comparisons made by the same person are dependent. This aspect has
received much attention in the literature during the last decade.

\subsubsection{Thurstonian models}\label{secthurstone} The original model\break proposed by
\citet{Thurstone27} includes correlation\break among the observations. The
model was developed for analyzing sensorial discrimination and assumes
that the stimuli $\mathbf{T}=(T_{1}, \ldots, T_{n})^{\prime}$
compared in a~pair\-ed comparison experiment follow a normal
distribution, $\mathbf{T} \sim N(\bolds{\mu}, \bolds{\Sigma}_{T})$,
with mean $\bolds{\mu}=(\mu_{1},
\dots,\mu_{n})^{\prime}$ and variance $\bolds{\Sigma}_{T}$.
\citet{Thurstone27} proposes different models with different
covariance matrices of the stimuli, so the set of models which assume a
normal distribution of the stimuli are called Thurstonian models. The
single realization $t_{i}$ of the stimulus $T_{i}$ can vary, and the
result of the paired comparison between the same two stimuli can be
different in different occasions. Assume that only either a preference
for $i$ or a preference for $j$ can be expressed, so then in a paired
comparison when $T_{i}>T_{j}$ object $i$ is preferred, or
alternatively, when the latent random variable $Z_{ij}=T_{i}-T_{j}$
is
positive, a win for $i$ is observed;\vadjust{\goodbreak} otherwise a win for $j$ occurs. In
the context of multiple judgment sampling, \citet{Takane89} proposes
to include a vector of pair specific errors. Let $\mathbf{Z}_{s}=(Z_{s
12}, \ldots, Z_{s\,n-1\,n})^{\prime}$ be the vector of
all latent continuous random variables pertaining to subject~$s$, then
%
%
\begin{equation}\label{eqtak}
\mathbf{Z}_{s}= \mathbf{A} \mathbf{T}+ \mathbf{e}_{s},
\end{equation}
where $\mathbf{e}_{s}=(e_{s12}, e_{s13}, \ldots, e_{s\,n-1\,
n})^{\prime}$ is the vector of pair-specific errors which has zero
mean, covariance $\bolds{\Omega}$ and is independent of
$\mathbf{T}$ and of $\mathbf{e}_{s^{\prime}}$ for any other
subject $s^{\prime} \neq s$, and $\mathbf{A}$ is the design matrix of paired
comparisons whose rows identify the paired comparisons and columns
correspond to the objects. For example, if $n=4$, and the paired
comparisons are $(1,2), (1,3), (1,4), (2,3), (2,4)$ and $(3,4)$, then
\[
\mathbf{A}=
\pmatrix{
1 & -1 & 0 & 0\cr
1 & 0 & -1 & 0\cr
1 & 0 & 0 & -1\cr
0 & 1 & -1 & 0\cr
0 & 1 & 0 & -1\cr
0 & 0 & 1 & -1\cr
}.
\]
A similar model is employed by \citet{BockenholtTsai01}, who assume
that $\bolds{\epsilon}_{s} \sim N(\mathbf{0}, \omega^{2}
\mathbf{I}_{N})$. The more general analysis of covariance structure
proposed by \citet{Takane89} can accommodate both the {\em wandering
vector} and the {\em wandering ideal point} models (\cite{Carroll91}),
which are models with different assumptions about the mechanism
originating the data. The wandering vector and wandering ideal point
models do not impose the number of dimensions which is determined from
the data alone, so they are powerful models to analyze human choice
behavior and inferring perceptual dimensions.

The model thus specified is over-parametrized. To reduce the number of
parameters, \citet{Thurstone27} proposes different restrictions on the
covariance matrix $\bolds{\Sigma}_{T}$, while \citet{Takane89}
proposes a factor model. Nonetheless, these models with a reduced
number of parameters need further identification restrictions; see
Section \ref{sec522}.

A further extension of model (\ref{eqtak}) is proposed by
\citet{Tsai08} who unify \citet{Tsai06} with \citet{Takane89} to
obtain\break a~general class of models that can account simultaneously for
transitive choice behavior and systematic deviations from it. In this
case the latent variable is
%
%
\begin{equation}\label{TB08}
\mathbf{Z}_{s}=\mathbf{A} \mathbf{T}+\mathbf{B} \mathbf{V}_{s},
\end{equation}
where $\mathbf{V}_{s}=(V_{s1(2)}, V_{s1(3)}, \ldots, V_{s2(1)},
V_{s2(3)}, \ldots,\break V_{s\,n\,(n-1)})^{\prime}$ is a vector of zero mean
random effects designed\vadjust{\goodbreak} so as to capture the random variation in
judging an object when compared to another specific object, and $\mathbf{B}$
is a matrix with rows corresponding to the paired comparisons and
columns corresponding to the elements of $\mathbf{V}_{s}$, so, for
example, if $n=3$, $\mathbf{V}_{s}=(V_{s1(2)}, V_{s1(3)}, V_{s2(1)},
V_{s2(3)}, V_{s3(1)}, V_{s3(2)})^{\prime}$ and
\[
\mathbf{B}=
\pmatrix{
1 & 0 & -1 & 0 & 0 & 0\cr
0 & 1 & 0 & 0 & -1 & 0\cr
0 & 0 & 0 & 1 & 0 & -1\cr
}.
\]

It is assumed that $\mathbf{V}_{s}$, the within-judge variability,
is normally distributed with mean $0$ and covariance $\bolds{\Sigma
}_{V}$ so that $\mathbf{Z}_{s} \sim N(\mathbf{A} \bolds{\mu},
\mathbf{A} \bolds{\Sigma}_{T} \mathbf{A}^{\prime}+\mathbf{B}
\bolds{\Sigma}_{V} \mathbf{B}^{\prime})$.

In the remaining it will be assumed that there are only two possible
outcomes of the comparisons, but it is easy to extend this model for
ordinal data through the introduction of threshold parameters with a
specification analogous to (\ref{cumulativeModel}).

\subsubsection{Identification}
\label{sec522} Psychometricians are interest\-ed in understanding the
relations between stimuli; hence they are primarily interested in the
unstructured and unrestricted Thurstonian models. Unfortunately, due to
the comparative nature of the data, some identification restrictions on
the covariance matrix are needed. The necessary identification
restrictions to estimate model (\ref{eqtak}) are discussed in
May\-deu-Olivares (\citeyear{Maydeu01}, \citeyear{Maydeu03}), Tsai and B{\"o}ckenholt\break
(\citeyear{Tsai02}) and
\citet{Tsai03}. Consider the covariance matrix $\bolds{\Sigma
}_{Z}=\operatorname{Cov}(\mathbf{Z}_{s})=\mathbf{A} \bolds{\Sigma}_{T} \mathbf{A}^{\prime}+\bolds{\Omega}$, where
$\bolds{\Sigma}_{ T}$ is an unrestricted covariance matrix. Because
of the difference structure of the judgments $\bolds{\Sigma}_{ T}$
and $\bolds{\Sigma}_{ T}+ \mathbf{d} \mathbf{1}^{\prime}+
\mathbf{1} \mathbf{d}^{\prime}$ where $\mathbf{1}$ is a~vector
of $n$ ones and $\mathbf{d}$ is an $n$-dimensional vector of
constants such that the matrix remains positive definite, are not
distinguishable (\cite{Tsai00}). Indeed, let $\mathbf{K}=[ \mathbf{I}_{n-1} |
{-}\mathbf{1}]$ be an identity matrix of dimension $n-1$ to
which a column of elements equal to $-1$ is added, then only~$\mathbf{K}
\bolds{\mu}$ and $\mathbf{K} \bolds{\Sigma}_{T} \mathbf{K}^{\prime}$ are
identifiable. For example the matrices
\begin{eqnarray*}
\bolds{\Sigma}_{T,1} &=&
\pmatrix{
1 & 0 & 0\cr
0 & 1 & 0\cr
0 & 0 & 1\cr
}, \\ \bolds{\Sigma}_{T,2} &=&
\pmatrix{
0.750 & 0.125 & 0\cr
0.125 & 1.5 & 0.375\cr
0 & 0.375 & 1.250
}
\end{eqnarray*}
are not distinguishable because $ \mathbf{K} \bolds{\Sigma}_{
T,\mathrm{1}} \mathbf{K}^{\prime}=\mathbf{K} \cdot\break\bolds{\Sigma}_{
T,\mathrm{2}} \mathbf{K}^{\prime}= \bigl({{2 \atop 1} \enskip {1
\atop 2}}\bigr)$, where the second matrix is obtain\-ed from
the first one by setting $\mathbf{d}=(-1/8, 1/4, 1/8)$. This
consideration remains valid for any generic matrix of contrasts that
may be used instead of $\mathbf{K}$. The specifications of the covariance
matrix $\bolds{\Sigma}_{T}$ with a~reduced number\vadjust{\goodbreak} of parameters
proposed by \citet{Thurstone27} cannot be recovered from the data and
only covariance classes can be considered.

\citet{Tsai03} shows that $n+2$ constraints are needed in order to
identify model (\ref{eqtak}), including the constraint on the worth
parameters. As for the mean parameters, many different constraints can
be imposed on the covariance matrix. For example,
\citet{BockenholtTsai01}, \citet{Tsai02} and \citet
{Maydeu03} set
all the diagonal elements of $\bolds{\Sigma}_{T}$ equal to 1 and
either one of the diagonal elements of $\bolds{\Omega}$ to 1 or one
of the nondiagonal elements of $\bolds{\Sigma}_{T}$ equal to zero.
However, if $\bolds{\Sigma}_{T}$ is fixed to be a~correlation
matrix, the set of matrixes that produce the same sets of probabilities
is limited. \citet{Maydeu05} set all the covariances involving the
last latent utility to zero, which corresponds to assuming independence
between the last stimulus and the others, and the variance of the first
and last item to one. Maydeu-Oliva\-res (\citeyear{Maydeu07}) suggest to set all diagonal
elements of~$\bolds{\Sigma}_{T}$ equal to one and the sum of the
correlations between the first and the other latent variables to one.
With these constraints positive entries in the correlation matrix imply
that strong preference for one stimulus is associated with strong
preference for the other stimulus, while negative entries indicate that
strong preference for one stimulus is associated with weak preference
for the other stimulus. Thus, it is not necessary to fix any element in
the matrix $\bolds{\Omega}$, since the constraint $\omega=1$ in
$\bolds{\Omega}= \omega^{2} \mathbf{I}_{N}$ could lead to a
nonpositive definite matrix $\bolds{\Sigma}_{T}$. After estimation
it is possible to recover the class of covariance matrixes that produce
the same probabilities (\cite{Maydeu07}). However, the initial
identification constraints pose limits on the set of covariance
matrixes that identify the same model.

There is no discussion or results about the identification restrictions
necessary to estimate model~(\ref{TB08}). In order not to incur
identification problems, \citet{Tsai08} assume that the
matrix~$\bolds{\Sigma}_{V}$ depends on very few parameters.

\subsubsection{Models with logit link}

The dependence between evaluations made by the same judge has been
introduced also in models employing logit link functions. Different
specifications have been used for this purpose.

A first inclusion of dependence in logit models is proposed by
\citet{Lancaster83}, who consider multiple judgments by the same
person and introduce correlation in the Bradley--Terry model assuming
that the worth parameters are random variables\vadjust{\goodbreak} following a beta
distribution with shape parameters $a_{ij}$ and $b_{ij}$. The
Bradley--Terry model is imposed on the means of the beta distributions,
that is, $\mathrm{E}(\pi_{ij})=a_{ij}/(a_{ij}+b_{ij})=\pi_{i}/(\pi
_{i}+\pi_{j})$, but such
a model introduces correlation only between comparisons of the same
judge on the same pair of objects, while the other comparisons remain
independent. The same limit presents the extension by
\citet{Matthews95} who consider three possible response categories.

Two different methods have been used for introducing dependence among
comparisons made by the same person involving one common object in
logit models. The first method exploits the usual association measure
for binary data: the odds ratio. B{\"o}cken\-holt and Dillon (\citeyear{Bockenholt97}) consider the
adjacent categories model for preference data with $H$ categories and
suggest a parametrisation in terms of log-odds ratios to account for
dependence between observations, while \citet{Dittrich02} adopt a
similar approach in a two-categorical model using the log-linear
formulation of the Brad\-ley--Terry model. This specification is
convenient because it allows one to estimate the model through standard
software developed for log-linear models, but the number of added
parameters can be quite large (\cite{Dittrich02}).

Another method used for introducing dependence among observations is
the inclusion of random effects in the linear predictor.
\citet{Bockenholt01} describes the worth of object $i$ for subject $s$
as
\[
\mu_{si}=\mu_{i}+\sum_{p=1}^{P}\beta_{ip} x_{ip}+U_{si},
\]
where $U_{si}$ is a random component, and $\mathbf{x}_{i}$ is a
vector of $P$ subject-specific (and possibly item specific) covariates.
\citet{Bockenholt01} employs a logit link function and assumes that
$\mathbf{U}_{s}=(U_{s1}, \ldots, U_{sn})^{\prime}$ follows a
multivariate normal distribution with mean~$\mathbf{0}$ and
covariance~$\bolds{\Sigma}_{U}$.

\citet{Francis10} consider the log-linear representation of the
Bradley--Terry model and introduce random effects for each respondent
in order to account for residual heterogeneity that is not included in
subject-specific covariates. The inclusion of random effects in the
linear predictor introduces difficulties in the estimation of the
model.

\subsubsection{Choice models}
The work by Thurstone has great importance in the development of models
for analyzing discrete choices,\vadjust{\goodbreak} not only from a psychometric point of
view, but also in economic choice theory. When the idea that choices
may be random and not fixed started to develop, the use of the model
proposed by Thurstone was suggested (\cite{Marschak60}). As the Nobel
laureate \citet{McFadden01} states, ``when the perceived stimuli are
interpreted as levels of satisfaction, or utility, this can be
interpreted as a model for economic choice.''

According to the economic theory, models for discrete choice are
required to satisfy the utility maximization assumption which states
that subjects maximize their utility when making decisions. Let
$\Upsilon_{si}$ denote the utility of subject $s$ from alternative $i$
which can be decomposed as $\Upsilon_{si}=M_{si}+\epsilon_{si}$,
where $M_{si}$ denotes a function which relates a set of alternative
attributes and subject attributes to the utility gain and
$\epsilon_{si}$ denotes factors that affect utility, but are not
included in $M_{si}$. The probability that subject $s$ chooses
alternative $i$ is equal to the probability that the utility gained
from $i$ is higher than the utility from every other object in the
choice set: $\operatorname{pr}(\Upsilon_{si}> \Upsilon_{sj}, \forall
i \neq
j)=\operatorname{pr}(\epsilon_{si}-\epsilon_{sj} < M_{si}-M_{sj},\break
\forall  i
\neq j)$. These models are called random utility models. For each
person, a choice is described as $n-1$ paired comparisons between the
preferred alternative and all other options. Note that paired
comparisons do not really occur, so inconsistent choices cannot be
observed.

From the above specification, different models have been developed
depending on the assumptions about the distribution of the errors and
the formulation of the mean term $M_{si}$. If the $\epsilon_{si}$'s are
independent and follow a Gumbel distribution the choice model is
a~logit model and, when $M_{si}=\mathbf{x}_{si}^{\prime} \bolds{\beta}$,
it corresponds to the structured Bradley--Terry model. A
particular concern is caused by the independence from irrelevant
alternatives (\cite{Luce59}) property which characterizes the
Bradley--Terry model. Indeed, in the Bradley--Terry model the ratio
between probabilities of choosing one option over another is
independent from the other available alternatives. Often, this property
is not satisfied in real data. This limit is somehow overcome by
assuming a type of generalized extreme value distribution for the
errors. In the resulting nested logistic model, independence from
irrelevant alternatives holds for sets of alternatives within a same
subset and not for alternatives in different subsets (\cite{Train09}).
The advantage of these specifications is that models can be estimated
easily, but they cannot account for random taste variation or
unobserved factors correlated over time.\looseness=-1

A further proposal is to assume a multivariate normal distribution for
the errors $\epsilon_{si}$. This model is very flexible since it allows
for random taste variation and, when necessary, for temporally
correlated errors, but its estimation is not straightforward. The
resulting model is a multivariate probit model, like the Thurstone
model. In economic choice models it is of interest to consider the
influence on decisions of covariates that are included in the mean term~$M_{si}$.
Explanatory variables can be considered also in psychometric
models (\cite{Tsai02}), even though interest is focused on the
parameters $\bolds{\mu}$ which are always included in the linear
predictor.

Other extensions include further random elements in the mean term
$M_{si}$, so as to allow flexible disturbances or to account for
different attitudes and perceptions of different people. All these
elements add difficulties in the estimation of the model.

An important aspect in choice theory is the distinction between stated
and revealed preferences. This problem has not received much attention
in the psychometric literature, but there may be differences between
what people say they would choose in a questionnaire survey and what
they really choose. The former are called stated preferences and the
latter revealed preferences. If both types of preferences are
available, it may be useful to analyze them all together.
\citet{Walker02} propose a model that incorporates many of the above
extensions; however, care is needed when specifying the model because
it may be difficult to understand which parameters can be identified.
Moreover, the inclusion of additional disturbances and unobserved
covariates requires the approximation of integrals whose dimension can
be high.

Random utility models are very useful and widely spread; however, some
doubts have been raised about their basic assumption that people act as
to maximize their utility since sometimes consumers do not make
rational choices (\cite{Bockenholt06}).

\subsection{Object-Related Dependencies}\label{sec52}
In the multiple judgment sampling the dependence among
observations derives from repeated comparisons made by the same person,
usually involving a~common object. In case paired comparisons are not
performed by a judge, the correlation may arise from the fact that the
same object is involved in multiple paired comparisons. For example,
when contests among animals are analyzed, it is realistic to assume\vadjust{\goodbreak}
that comparisons involving the same animal are correlated. In this
perspective, \citet{Firth05} suggests to set
%
%
\begin{equation}\label{eqre}
\mu_{i}=\mathbf{x}_{i}^{\prime} \bolds{\beta}
+U_{i},
\end{equation}
where $U_{i}$ is a zero mean object-specific random effect. This
approach is investigated in \citet{Cattelan09}.
The results of comparisons are related to observed characteristics of
the animal and to unobserved quantities that are captured by the random
effect $U_{i}$.

In this case, the latent random variable can be written as
\[
\mathbf{Z}=\mathbf{A X} \bolds{\beta}+\mathbf{A}\mathbf{U} + \bolds{\eta},
\]
where $\mathbf{U}=(U_{1}, \ldots, U_{n})$ is the vector of all
object-specific random effects, $\mathbf{X}$ is the matrix of
covariates with columns $\mathbf{x}_{i}$, $\bolds{\eta}$ are
independent normally distributed errors with mean 0 and variance 1
while the matrix $\mathbf{A}$ is the design matrix of the paired
comparisons with rows that describe which comparisons are observed, not
necessarily all possible paired comparisons. If it is assumed that
$\mathbf{U}$ is multivariate normal with mean $\mathbf{0}$ and
covariance $\mathbf{I}_{n} \sigma^{2}$, then $ \mathbf{Z} \sim
N(\mathbf{A X} \bolds{\beta}, \sigma^{2} \mathbf{AA}^{T}+\mathbf{I}_{d})$,
where $d$ is the number of paired
comparisons observed. Again, this model is a multivariate probit model.
However, this type of data presents some different features with
respect to multiple judgment sampling.
While in pshychometric applications $n$ is not very large because it is
unlikely that a person will make all the paired comparisons when
$n>10$, this will typically happen in sport tournaments or in paired
comparison data about animal behavior. Moreover, in the multiple
judgment sampling scheme $S$ independent replications of all the
comparisons are available, but in other contexts this does not occur,
adding further difficulties.

\subsection{Inference}

\subsubsection{Estimation}\label{secestimation}
In this section, the multiple judgment sampling
scheme is mainly investigated, and only some comments are made about
the case of object-related dependencies. There are different methods
for estimating models for dependent paired comparison data. A first
approach to the computation of the likelihood function requires to
integrate out the latent variables $\mathbf{T}$ from the joint
distribution of $\mathbf{Y}$ and $\mathbf{T}$. This integral has
dimension $n$, the number of items, but rewriting it in terms of
differences $T_{i}-T_{n}$, $i=1, \ldots, n-1$, the dimension can be
reduced to $n-1$, which nonetheless may still be quite large when
methods such as the Gauss--Hermite quadrature are employed.\vadjust{\goodbreak}

Alternatively, it is possible to represent the joint distribution of
the observations as a multivariate probit model. Let $\mathbf
{Z}_{s}^{*}=\mathbf{D}(\mathbf{Z}_{s}-\mathbf{A} \bolds{\mu})$ be
the standardized version of the latent variable $\mathbf{Z}_{s}$,
where $\mathbf{D}=[\operatorname{diag}(\bolds{\Sigma}_{
Z})]^{-1/2}$ and $\bolds{\Sigma}_{Z}$ denotes the
covariance matrix of $\mathbf{Z}_{s}$ expressed as in model
(\ref{eqtak}) or in model (\ref{TB08}). Then, $\mathbf{Z}_{s}^{*}$
follows a multivariate normal distribution with mean $\mathbf{0}$
and correlation matrix $\bolds{\Sigma}_{Z^{*}}=\mathbf{D} \bolds{\Sigma}_{
Z} \mathbf{D}$. Object $i$ is preferred to object $j$
when $z^{*}_{sij} \geq\tau^{*}_{ij}$, where the vector of the
thresholds is given by $\bolds{\tau}^{*}=-\mathbf{D}\mathbf{A}\bolds{\mu}$. The
likelihood function is the product of the probability of the
observations for each judge
\[
\mathcal{L}(\bolds{\psi}; \mathbf{Y})=
\prod_{s=1}^{S}\mathcal{L}_{s}(\bolds{\psi}; \mathbf{Y}_{s}),
\]
where
\[
\mathcal{L}_{s}(\bolds{\psi}; \mathbf{Y}_{s})=\int_{R_{s12}}
\cdots\int_{R_{s n-1   n}} \phi_{N}(\mathbf{z}_{s}^{*};
\bolds{\Sigma}_{Z^{*}})\,\mathrm{d} \mathbf{z}_{s}^{*},
\]
$\phi_{N}(\cdot; \bolds{\Sigma}_{Z^{*}})$ denotes the density
function of an $N$-dimensional normal random variable with mean
$\mathbf{0}$ and correlation matrix $\bolds{\Sigma}_{Z^{*}}$ and
\[
R_{sij}=
\cases{
(- \infty, \tau^{*}_{ij}) & $\mbox{if } Y_{sij}=1$,\cr
(\tau^{*}_{ij}, \infty) & $\mbox{if } Y_{sij}=2$.
}
\]
Note that this approach requires the approximation of $S$ integrals
whose dimension is equal to $N=n   (n-1)/2$, the number of paired
comparisons, so its growth is quadratic with the increase in the number
of objects. However, there is a large literature about methods for
approximate inference in multivariate probit models. The algorithm
proposed by \citet{Genz02} to approximate multivariate normal
probabilities is based on quasi-Monte Carlo methods, and
\citet{Craig08} warns against the randomness of this method for
likelihood evaluation. A~deterministic approximation is developed by
\citet{Miwa03}, but it is available only for integrals of
dimension up
to 20 since even for such a dimension its computation is very slow.
Approximations based on Monte Carlo methods can be used
(\cite{Chib98}), but they may be computationally expensive if the
dimension of the integral is very large. \citet{BockenholtTsai01} use
an EM algorithm, while in econometric theory a maximum simulated
likelihood approach in which multivariate normal probabilities are
simulated through the Geweke--Hajivassiliou--Keane algorithm is
employed (\cite{Train09}). A further approach may be based on data
cloning (\cite{Lele10}). When integrals are very large, and the
approximation is\vadjust{\goodbreak} computationally demanding and time-consuming, it is
possible to resort to limited information estimation methods, which are
estimation procedures based on low dimensional margins. Here, we
compare two different methods. The first one is widely applied in the
context of multiple judgment sampling (\citeauthor{Maydeu01}, \citeyear{Maydeu01,Maydeu02};
Maydeu-\break Olivares and B{\"o}ckenholt \citeyear{Maydeu05})
and will be called limited information estimation; the
second is proposed in the context of object-specific dependencies in
\citet{Cattelan09} and is called pairwise likelihood.

The limited information estimation procedure considered here consists
of three stages. In the first stage the threshold parameters
$\bolds{\tau}^{*}$ are estimated exploiting the empirical univariate
proportions of wins. In the second stage the elements of $\bolds{\Sigma
}_{Z^{*}}$, which are tetrachoric correlations, are estimated
employing the bivariate proportions of wins. Finally, in the third
stage the model parameters $\bolds{\psi}$ are estimated by
minimizing the function
%
%
\begin{equation}\label{eqG}
G= \{\tilde{\bolds{\kappa}} - \bolds{\kappa}(\bolds{\psi})
\}^{\prime} \hat{\mathbf{W}}
\{\tilde{\bolds{\kappa}}-\bolds{\kappa}(\bolds{\psi}) \},
\end{equation}
where $\tilde{\bolds{\kappa}}$ denotes the thresholds, and
tetrachoric correlations, estimated in the first and second stages,
$\bolds{\kappa}(\bolds{\psi})$ denotes the thresholds, and
tetrachoric corre\-lations under the restrictions imposed on those
parameters by the model parameters $\bolds{\psi}$ and $\hat
{\mathbf{W}}$ is a~non\-negative definite matrix. Let $\bolds{\Xi}$
denote the asymptotic covariance matrix of $\tilde{\bolds{\kappa}}$.
Then it is possible to~use $\hat{\mathbf{W}}=\hat{\bolds{\Xi}}^{-1}$
(\cite{Muthen78}), $\hat{\mathbf{W}}=[\operatorname{diag}(\hat{\bolds{\Xi
}})]^{-1}$ (\cite{Muthen97}) or
$\hat{\mathbf{W}}=\mathbf{I}$\break (\cite{Muthen93}). The last two
options seem more stable in data sets with a small number of objects
(\cite{Maydeu01}). This method is very fast, and \citet{Maydeu01}
states that it may have an edge over full information methods because
it uses only the one and two-dimensional marginals of a large and
sparse contingency table.

Pairwise likelihood (\cite{LeCessie94}) is a special case of the
broader class of composite likelihoods (\cite{Lindsay88}; \cite{Varin11}).
The pairwise likelihood of all the observations is the product of the
pairwise likelihoods relative to the single judges $\mathcal{L}_{
\mathrm{pair}}(\bolds{\psi}; \mathbf{Y})=\prod_{s=1}^{S}
\mathcal{L}^{s}_{\mathrm{pair}}(\bolds{\psi}; \mathbf{Y}_{s})$,
where
\begin{eqnarray*}
&&\mathcal{L}^{s}_{\mathrm{pair}}(\bolds{\psi}; \mathbf{Y}_{s})\\
&&\quad=\prod
_{i=1}^{n-2} \prod_{j=i+1}^{n-1} \prod_{k=i}^{n-1}
\prod_{l=j+1}^{n}\operatorname{pr}(Y_{sij}=y_{sij}, Y_{skl}=y_{skl}).
\end{eqnarray*}

%
\begin{table*}[b]
\tabcolsep=0pt
\caption{Average (\texttt{Mn}) and median (\texttt{Md}) simulated
estimates, average model-based standard errors (\texttt{s.e.}) and
simulation~standard~deviations (\texttt{s.d.}) of parameters estimated
by maximum likelihood (\texttt{ML}),
limited~information~estimation~(\texttt{LI})~and pairwise likelihood (\texttt{PL})}\label{tabSim1}
\begin{tabular*}{\textwidth}{@{\extracolsep{4in minus 4 in}}ld{2.1}d{2.2}ccd{2.2}d{2.2}ccd{2.2}d{2.2}cc@{}}
\hline
 &\multicolumn{1}{c}{\multirow{2}{20pt}[-7pt]{\centering\textbf{True value}}}& \multicolumn{3}{c}{\textbf{ML}} &  \multicolumn{4}{c}{\textbf{LI}}
& \multicolumn{4}{c@{}}{\textbf{PL}} \\
\ccline{3-5,6-9,10-13}\\ [-7pt]
 && \multicolumn{1}{c}{\textbf{Mn}} & \multicolumn{1}{c}{\textbf{s.e.}}
& \multicolumn{1}{c}{\textbf{s.d.}} & \multicolumn{1}{c}{\textbf{Mn}} & \multicolumn{1}{c}{\textbf{Md}}
& \multicolumn{1}{c}{\textbf{s.e.}} & \multicolumn{1}{c}{\textbf{s.d.}}
& \multicolumn{1}{c}{\textbf{Mn}} & \multicolumn{1}{c}{\textbf{Md}} & \multicolumn{1}{c}{\textbf{s.e.}} &
\multicolumn{1}{c@{}}{\textbf{s.d.}}\\
\hline
$\mu_{1}$& 0.5 & 0.51 & 0.13 & 0.13 & 0.51 & 0.50 &
0.13 & 0.13
& 0.50 & 0.50 & 0.13 & 0.13 \\
$\mu_{2}$& 0 & 0.01 & 0.12 & 0.13 & 0.01 & 0.01 &
0.12 & 0.13& 0.01 & 0.01 & 0.12 & 0.13\\
$\mu_{3}$& -0.5 & -0.49 & 0.15 & 0.15 & -0.50 & -0.48
& 0.15 & 0.15 & -0.49 & -0.48 & 0.15 & 0.15\\
$\sigma_{12}$& 0.8 & 0.80 & 0.12 & 0.14 & 0.78 & 0.80 &
0.13 & 0.14 & 0.79 & 0.80 & 0.13 & 0.15\\
$\sigma_{13}$& 0.7 & 0.70 & 0.17 & 0.17 & 0.69 & 0.71 &
0.17 & 0.17 & 0.69 & 0.71 & 0.18 & 0.18\\
$\sigma_{14}$& 0.8 & 0.79 & 0.13 & 0.14 & 0.78 & 0.79 &
0.13 & 0.14 & 0.78 & 0.80 & 0.14 & 0.15\\
$\sigma_{23}$& 0.6 & 0.58 & 0.19 & 0.20 & 0.57 & 0.60 &
0.19 & 0.20 & 0.57 & 0.60 & 0.19 & 0.20\\
$\sigma_{24}$& 0.7 & 0.68 & 0.16 & 0.16 & 0.66 & 0.67 &
0.16 & 0.17 & 0.67 & 0.68 & 0.16 & 0.17 \\
$\sigma_{34}$& 0.6 & 0.58 & 0.21 & 0.20 & 0.57 & 0.60 &
0.20 & 0.20 & 0.57 & 0.60 & 0.20 & 0.20\\
\hline
\end{tabular*}
\end{table*}

Let $\ell^{s}_{\mathrm{pair}}(\bolds{\psi}; \mathbf{Y}_{s})=\log
\mathcal{L}^{s}_{\mathrm{pair}}(\bolds{\psi}; \mathbf{Y}_{s})$
denote the logarithm of the pairwise likelihood for subject $s$ and
$\ell_{\mathrm{pair}}(\bolds{\psi}; \mathbf{Y})=\sum_{s=1}^{S}\ell
^{s}_{\mathrm{pair}}(\bolds{\psi}; \mathbf{Y}_{s})$ be the whole pairwise
log-likelihood. Under usual regularity
conditions on the log-likelihood of univariate and bivariate margins,
the maximum pairwise likelihood estimator is consistent and
asymptotically normally distributed with mean $\bolds{\psi}$ and
covariance matrix\break $\mathbf H(\bolds{\psi})^{-1} \mathbf
J(\bolds{\psi}) \mathbf H(\bolds{\psi})^{-1}$, where $\mathbf
J(\bolds{\psi})=\mbox{var} \{ \nabla
\ell_{\mathrm{pair}}(\bolds{\psi};\break \mathbf{Y}) \}$ and $\mathbf
H(\bolds{\psi})= E \{ -\nabla^{2}\ell_{\mathrm{pair}}(\bolds{\psi}; \mathbf{Y}) \}$
(\cite{Molenberghs05}; \cite{Varin11}). Unfortunately, the analogous of the
likelihood ratio test based on pairwise likelihood does not follow the
usual chi-square distribution (\cite{Kent82}). In the multiple judgment
sampling context, it is natural to consider asymptotic properties of
pairwise likelihood estimators computed as the number of subjects
increases, that is, as $S \rightarrow\infty$. When the number of paired
comparisons per subject is bounded, the above properties are satisfied
(\cite{Zhao05}). Pairwise likelihood reduces noticeably the
computational effort since it requires only the computation of
bivariate normal probabilities. The standard errors can be computed
straightforwardly by exploiting the independence between the
observations of different judges. In fact, $\mathbf H(\bolds{\psi})$
can be estimated by the Hessian matrix computed at the maximum
pairwise likelihood estimate, while the cross-product $\sum_{s=1}^{S}
\nabla\ell^{s}_{ \mathrm{pair}}(\hat{\bolds{\psi}}; \mathbf{Y}_{s})
\nabla\ell^{s}_{\mathrm{pair}}(\hat{\bolds{\psi}}; \mathbf{Y}_{s})^{\prime
}$ can\vspace*{1pt} be used to estimate $\mathbf J(\bolds{\psi})$.

The case of object-related dependencies is not considered in the
following simulation study; however, note that some different
difficulties arise. As already pointed out, in this context there is a
large $n$ and small $S$, so the limited information estimation meth\-od
cannot be applied, but pairwise likelihood can still be
employed
(\cite{Cattelan09}). However, it is more problematic to consider the
asymptotic behavior of the maximum pairwise likelihood estimator when
data are a long sequence of dependent observations; see, for example,
\citet{Cox}. In the context of paired comparison data, results of
simulations for increasing $n$ when all possible paired comparisons are
performed are encouraging (\cite{Cattelan09}); however, theoretical
results for this instance are still lacking.

\subsubsection{Simulation studies}
Simulation studies were performed considering models (\ref{eqtak}) and
(\ref{TB08}). It is assumed that $n=4$; hence also a full likelihood
approach based on the algorithm by \citet{Miwa03} can be used since
the integral has dimension $6$.

The first simulation setting is the same as that proposed in
\citet{Maydeu01}, where the mod\-el $\mathbf{Z}_{s} = \mathbf{A}
\mathbf{T} + \mathbf{e}_{s}$ is assumed with
\[
\bolds{\mu}=
\pmatrix{
0.5\cr
0\cr
-0.5\cr
0\cr
},
\quad
\bolds{\Sigma}_{T}=
\pmatrix{
1 & & & \cr
0.8 & 1 & & \cr
0.7 & 0.6 & 1 & \cr
0.8 & 0.7 & 0.6 & 1\cr
}
\]
and the covariance matrix of $\mathbf{e}$ is $\bolds{\Omega}=\omega^{2}
\mathbf{I}_{6}$. For identification purposes the
diagonal elements of $\bolds{\Sigma}_{T}$ are set equal to 1,
$\mu_{4}=0$ and $\omega^{2}=1$. Hence, in this case~$\bolds{\Sigma
}_{T}$ is actually a correlation matrix. Table \ref{tabSim1}%
%
\begin{table}
\tabcolsep=0pt
\caption{Empirical coverage of confidence intervals for model
parameters of limited information estimator (\texttt{LI}) and~pairwise
likelihood estimator (\texttt{PL}) at nominal~levels~$95\%$,~$97.5\%$~and $99\%$}\label{tabcoverage1}
\begin{tabular*}{\columnwidth}{@{\extracolsep{4in minus4in}}lcccccc@{}}
\hline
& \multicolumn{2}{c}{\textbf{0.950}} & \multicolumn{2}{c}{\textbf{0.975}} &
\multicolumn{2}{c@{}}{\textbf{0.990}}\\
\ccline{2-3,4-5,6-7}\\[-7pt]
& \textbf{LI} & \textbf{PL} & \textbf{LI} & \textbf{PL} & \textbf{LI} & \textbf{PL}\\
\hline
$\mu_{1}$ & 0.947 & 0.958 & 0.982 & 0.978 & 0.992 & 0.992\\
$\mu_{2}$ & 0.960 & 0.964 & 0.978 & 0.976 & 0.988 &0.988\\
$\mu_{3}$ & 0.941 & 0.930 & 0.969 & 0.972 & 0.995 &0.991\\
$\sigma_{12}$& 0.959 & 0.985 & 0.975 & 0.997 & 0.989 & 1.000\\
$\sigma_{13}$& 0.934 & 0.939 & 0.961 & 0.967 & 0.968 &0.985\\
$\sigma_{14}$& 0.941 & 0.968 & 0.967 & 0.996 & 0.988 &1.000\\
$\sigma_{23}$& 0.965 & 0.970 & 0.973 & 0.980 & 0.987 & 0.995\\
$\sigma_{24}$& 0.943 & 0.933 & 0.951 & 0.959 & 0.967 &0.973\\
$\sigma_{34}$& 0.953 & 0.946 & 0.969 & 0.966 & 0.977 &0.989\\
\hline
\end{tabular*}
\vspace*{-3pt}
\end{table}
%
%
%
%
%
\begin{table*}[b]
\tabcolsep=0pt
\caption{Average (\texttt{Mn}) and median (\texttt{Md}) simulated
estimates, average model-based standard errors (\texttt{s.e.}) and
simulation~standard~deviations (\texttt{s.d.}) of parameters estimated
by maximum likelihood (\texttt{ML}), limited~information~estimation~(\texttt{LI})
and pairwise likelihood (\texttt{PL})}\label{tabSim2}
\begin{tabular*}{\textwidth}{@{\extracolsep{4in minus 4in}}ld{2.1}d{2.2}ccd{2.2}d{2.2}ccd{2.2}d{2.2}cc@{}}
\hline
&\multicolumn{1}{c}{\multirow{2}{20pt}[-7pt]{\centering\textbf{True value}}}& \multicolumn{3}{c}{\textbf{ML}} &  \multicolumn{4}{c}{\textbf{LI}}
& \multicolumn{4}{c@{}}{\textbf{PL}} \\
\ccline{3-5,6-9,10-13}\\ [-7pt]
&& \multicolumn{1}{c}{\textbf{Mn}} & \multicolumn{1}{c}{\textbf{s.e.}} & \multicolumn{1}{c}{\textbf{s.d.}}&
\multicolumn{1}{c}{\textbf{Mn}} & \multicolumn{1}{c}{\textbf{Md}} & \multicolumn{1}{c}{\textbf{s.e.}} & \multicolumn{1}{c}{\textbf{s.d.}}
& \multicolumn{1}{c}{\textbf{Mn}} & \multicolumn{1}{c}{\textbf{Md}} & \multicolumn{1}{c}{\textbf{s.e.}} & \multicolumn{1}{c@{}}{\textbf{s.d.}}\\
\hline
$\tilde\mu_{1}$ & -0.2 & -0.21 & 0.19 & 0.18
& -0.23 & -0.21 & 0.21 & 0.22& -0.22 & -0.20 & 0.19 & 0.19\\
$\tilde\mu_{2}$ & 1 & 1.00 & 0.30 & 0.31 &
1.07 & 1.07 & 0.42 & 0.47 & 1.03 & 1.00 & 0.33 & 0.33\\
$ \tilde\mu_{3}$ & -1.5 & -1.51 & 0.31 & 0.32 &-1.59 & -1.59 & 0.49 & 0.51 & -1.54 & -1.51 & 0.36 &
0.35\\ [3pt]
$\tilde\sigma^{2}_{1}$ & 1.5 & 1.53 & 0.83 & 0.81 & 2.06 & 1.58 & 1.97 & 1.64 & 1.70 & 1.44 & 1.05 & 0.95\\[3pt]
$\tilde\sigma^{2}_{2}$ & 4 & 3.98 & 1.73 & 1.75 &5.34 & 4.37 & 4.42 & 4.48 & 4.45 & 3.92 & 2.42 & 2.15\\[3pt]
$\tilde\sigma^{2}_{3}$ & 3 & 3.01 & 1.41 & 1.42 &3.91 & 3.19 & 3.17 & 3.25 & 3.32 & 3.04 & 1.93 & 1.73\\[3pt]
$\tilde\sigma_{12}$ & 1 & 0.98 & 0.70 & 0.64 &1.34 & 1.06 & 1.44 & 1.30 & 1.12 & 0.97 & 0.87 & 0.77\\
$\tilde\sigma_{13}$ & 1.3 & 1.29 & 0.73 & 0.71 &1.72 & 1.39 & 1.48 & 1.49 & 1.43 & 1.27 & 0.95 & 0.84\\
$\tilde\sigma_{23}$ & 2.5 & 2.49 & 1.09 & 1.09 &3.35 & 2.72 & 2.67 & 2.77 & 2.77 & 2.48 & 1.53 & 1.33\\
$b$ & 0.5 & 0.53 & 0.41 & 0.39 &0.72 & 0.58 & 0.82 & 0.98 & 0.58 & 0.50 & 0.50 & 0.51\\
\hline
\end{tabular*}
\end{table*}
shows
the mean and medians of the simulated estimates on 1000 data sets
assuming $S=100$ judges. Moreover, the average of model-based standard
errors and the simulation standard deviations are reported. In limited
information estimation, the matrix $\hat{\mathbf{W}}=\mathbf{I}$ is
employed. In this setting all the methods seem to perform comparably
well. Table \ref{tabcoverage1} shows the empirical coverages of
confidence intervals based on the normal approximation.\vadjust{\goodbreak}

The second simulation setting considers model (\ref{TB08}) proposed by
\citet{Tsai08}. Here, we consider differences with a reference object,
so we compute means and variances of the differences $\tilde
T_{i}=T_{i}-T_{n}$ for $i=1, \ldots, n-1$. The assumed worth parameters
of these differences are $\tilde{\bolds{\mu}}=(-0.2, 1, -1.5)$
while the covariance matrix is
\[
\pmatrix{
1.5 & 1 & 1.3\cr
1 & 4 & 2.5\cr
1.3 & 2.5 & 3
},
\]
and $\tilde\sigma_{ij}$ is used to denote the element in row $i$ and
column $j$ of the above reduced matrix. Differently from the
previous\vadjust{\goodbreak}
setting, this specification of the model allows one to estimate also
the variance of the differences $T_{i}-T_{n}$ and to check whether they
are different for the various objects. \citet{Tsai08} propose a
specification of the matrix $\mathbf{B}$ which depends only on one
parameter $b$ whose value is set equal to 0.5.

Table \ref{tabSim2} presents the results of the simulations. Maximum
likelihood based on numerical integration is the method that performs
best; however, maximization of the likelihood was not always
straightforward, and sometimes the optimization algorithms employed
stopped at a point where the Hessian matrix was not negative definite.

Pairwise likelihood estimation seems to perform quite well, especially
if compared to limited information estimation, which seems not
satisfactory in this case with $S=100$, as already noticed in
\citet{Tsai08}. Estimating the parameters of the covariance matrix
appears more problematic than the estimation of the worth parameters,
and the average of the simulated estimates is particularly influenced
by some large values, but the median shows a better performance. In
particular, while the average simulated estimates for limited
information estimation shows a maximum percentage bias equal to
$44.1\%$, for the median it reduces to $15.4\%$. The maximum bias for
the mean of the simulated estimates using pairwise likelihood is
$16.1\%$, while for the median it is $4\%$. In both cases, pairwise
likelihood shows lower bias. The standard errors of pairwise likelihood
estimates are lower, thus yielding shorter confidence intervals. Table
\ref{tabcoverage2} reports the empirical coverage of Wald-type
confidence intervals for%
%
%
%
%
\begin{table}
\tabcolsep=0pt
\caption{Empirical coverage of confidence intervals for model
parameters of limited information estimator (\texttt{LI}) and~pairwise
likelihood estimator (\texttt{PL}) at nominal~levels~$95\%$,~$97.5\%$
and $99\%$}\label{tabcoverage2}
\begin{tabular*}{\columnwidth}{@{\extracolsep{4in minus 4in}}lcccccc@{}}
\hline
& \multicolumn{2}{c}{\textbf{0.950}} & \multicolumn{2}{c}{\textbf{0.975}} &
\multicolumn{2}{c}{\textbf{0.990}}\\
\ccline{2-3,4-5,6-7}\\[-7pt]
& \textbf{LI} & \textbf{PL} & \textbf{LI} & \textbf{PL} & \textbf{LI} & \textbf{PL}\\
\hline
$\tilde\mu_{1}$ & 0.955 & 0.935 & 0.981 & 0.965 & 0.994 & 0.983\\
$\tilde\mu_{2}$ & 0.962 & 0.960 & 0.973 & 0.974 & 0.986 & 0.986\\
$\tilde\mu_{3}$ & 0.920 & 0.938 & 0.941 & 0.960 & 0.961 & 0.977\\
$\tilde\sigma^{2}_{1}$& 0.932 & 0.922 & 0.947 & 0.936 & 0.959 &
0.966\\ [2pt]
$\tilde\sigma^{2}_{2}$& 0.932 & 0.924 & 0.949 & 0.945 & 0.964 &
0.961\\ [2pt]
$\tilde\sigma^{2}_{3}$& 0.936 & 0.937 & 0.949 & 0.953 & 0.963 &
0.964\\
$\tilde\sigma_{12}$& 0.932 & 0.937 & 0.951 & 0.956 & 0.966 & 0.970\\
$\tilde\sigma_{13}$& 0.915 & 0.912 & 0.929 & 0.933 & 0.939 & 0.945\\
$\tilde\sigma_{23}$& 0.922 & 0.920 & 0.941 & 0.937 & 0.953 & 0.951\\
$b$ & 0.936 & 0.936 & 0.946 & 0.953 & 0.963 &0.963\\
\hline
\end{tabular*}
\end{table}
%
%
%
%
%
%
\begin{table*}[b]
\caption{Estimates and standard errors (in brackets) of mean and
correlation parameters of model (\protect\ref{eqtak}) for universities data
using~constraints proposed by \protect\citet{Maydeu07}. In italics the
estimates and standard~errors~of~a~model with fixed correlation between~Paris and St. Gallen}\label{tabTU}
\begin{tabular*}{\textwidth}{@{\extracolsep{\fill}}lccccccc@{}}
\hline
& \textbf{Barcelona} & \textbf{London} & \textbf{Milan} & \textbf{Paris} & \textbf{St. Gallen} & \textbf{Stockholm} & $\bolds{\mu}$\\
\hline
Barcelona & 1 & \textit{$-$0.064} &\textit{0.688} &\textit{0.063} &\textit{$-$0.472} &
\textit{0.265} & 0.405 \\
& (fixed) & \textit{(0.183)} &\textit{(0.085)} &\textit{(0.158)} &\textit{(0.146)}
& \textit{(0.145)} & (0.073)\\
London & 0.058 & 1 &\textit{0.079} &\textit{$-$0.069} &\textit{$-$0.287} & \textit{
0.227} & 1.346 \\
& (0.084) & (fixed) & \textit{(0.185)} &\textit{(0.224)} &\textit{(0.147)} &
\textit{(0.154)} & (0.087)\\
Milan & 0.724 & 0.185 & 1 &\textit{0.244} &\textit{$-$0.466} & \textit{0.253} &
0.308 \\
& (0.062) & (0.097) & (fixed) &\textit{(0.174)} &\textit{(0.137)} & \textit{
(0.160)} & (0.074)\\
Paris & 0.171 & 0.054 & 0.331 & 1 &\textit{$-$0.690} & \textit{0.033} & 0.748 \\
& (0.094) & (0.117) & (0.113) & (fixed) & \textit{(fixed)} & \textit{
(0.267)} & (0.086)\\
St.\ Gallen & $-$0.303 & $-$0.139 & $-$0.298 & $-$0.496 & 1 &\textit{0.194} &
0.371 \\
& (0.113) & (0.139) & (0.144) & (0.157) & (fixed) & \textit{(0.135)} &
(0.081)\\
Stockholm & 0.350 & 0.316 & 0.339 & 0.144 & 0.287 & 1 & 0\\
& (0.079) & (0.091) & (0.097) & (0.113) & (0.130) & (fixed) & (fixed)\\
\hline
\end{tabular*}
\end{table*}
the estimated limited information estimation
and pairwise likelihood. The coverage rates of the two methods are very
similar, and in both cases the actual coverage for parameters of the
covariance matrix appears systematically lower than the nominal levels.
In order to obtain accurate coverage probabilities, we may need to
resort to a bootstrap procedure for detecting the distribution of the statistic,
while with pairwise likelihood it may be possible to obtain intervals
based on the pairwise likelihood function.

\begin{example*}
$\!\!\!$We fit model (\ref{eqtak}) to universities' data by
means of pairwise likelihood. A full likelihood approach based on
numeric approximation implies computing 303 integrals of dimension 5,
in case a~university is used as reference object, both for the mean and
covariance structure, but methods such as the Gauss--Hermite quadrature
are affected by the curse of dimensionality. A multivariate probit
approach would require a very slow computation because the algorithm by
Miwa would take very long to approximate 303 integrals of dimension 15.
It is assumed that $\bolds{\Omega}=\omega^{2} \mathbf{I}_{15}$.
Table \ref{tabTU} displays the results of the estimates, employing two
different sets of constraints. The lower triangle of the covariance
matrix shown in Table \ref{tabTU} reports the estimates obtained using
the constraints proposed in Maydeu-Olivares and Her\-n{\'a}ndez (\citeyear{Maydeu07}); see Section~\ref{sec522}. The estimate of the threshold parameter (with standard
error in brackets) is $\hat\tau_{2}=0.205\ (0.018)$ while the
variance parameter is $\hat\omega^{2}=0.180\ (0.026)$. A~high
correlation is estimated between Barcelona and Milan, so strong
preference for Barcelona is associated with strong preference for
Milan. Even though some correlations do not seem significant, it
appears that a strong preference for St. Gallen is associated with a
weak preference for all the other universities but Stockholm. The worth
parameters denote the same ranking of all universities as the one
arising from Table~\ref{tab1}. However, note that the estimated worth
parameters cannot be considered as absolute measures of worth of items;
indeed, it is possible to obtain alternative solutions that give an
equivalent fitting. The mean parameters that can be identified\vspace*{1.5pt} in the
model are standardized differences, that is,
$(\mu_{i}-\mu_{6})/\sqrt{\sigma^{2}_{i}+\sigma^{2}_{6}-2
\sigma_{i6}+\omega^{2}}$, $i=1, \ldots, 5$, where~$\mu_{6}$\break and~$\sigma^{2}_{6}$
are the mean and variance of the latent variable\vadjust{\goodbreak}
referring to Stockholm, the reference university. From the identified
parameters, different covariance matrixes of the universities can be
recovered. For example, in this instance where the matrix~$\bolds{\Sigma
}_{T}$ can be interpreted as a correlation matrix, it is shown
that the worth parameters $\sqrt{c}   \bolds{\mu}$, the
correlation matrix $c   \bolds{\Sigma}_{T} +(1-c) \mathbf{1}
\mathbf{1}^{\prime}$ and the covariance matrix of the pair-specific
errors $c   \bolds{\Omega}$ produce the same fitting of the model
for a positive constant $c$ such that the correlation matrix remains
positive definite (\cite{Maydeu07}). It is possible to set one of the
parameters of the correlation matrix according to some assumption, for
example we may presume that a strong preference for Paris is associated
with a weak preference for St. Gallen, and determine the value of $c$
which minimizes the correlation between the two universities while
yielding a positive definite correlation matrix. The value is $c=1.13$
which produces a correlation between Paris and St. Gallen equal to
$-$0.690. The estimates of the correlation matrix with this fixed value
of correlation between Paris and St. Gallen are shown in the upper
triangle of the matrix in Table \ref{tabTU}. The worth parameters can
be computed by multiplying the estimates shown in Table \ref{tabTU} by
$\sqrt{1.13}$. The fitting of the two models is equal, but in the
second case estimation is based on some previous theory about
correlation between a certain couple of universities.
\end{example*}

This analysis has only an illustrative purpose, in particular
\citet{Bockenholt01b} finds that a model with thresholds that vary
among subjects performs better than a model with a constant threshold
parameter.

%

\subsubsection{Model selection and goodness of fit}
\label{secmodel} Paired comparison data can be arranged in a
contingency table. In case of multiple judgment sampling the data can
be arranged in a table of dimension $2^{N}$ when there are two possible
outcomes and $H^{N}$ when the outcomes are $H$-categorical. As a
result, the contingency table will typically be very sparse, especially
if covariates are included so that paired comparisons are observed
conditional on the values of the covariates. In this situation the
likelihood ratio statistic and the Pearson statistic do not follow
a~$\chi^{2}$ distribution, nevertheless these statistics are often
employed to assess the model and for model selection. Differences
between observed and expected frequencies for subsets of the data, as
the $2 \times2$ subtables or triplets of comparisons, are sometimes
considered in order to identify where the fitting of the model is
not
good. In \citet{Dittrich07} the deviance is used\vadjust{\goodbreak} for selection between
nested models, but the test of goodness of fit cannot be based on the
asymptotic $\chi^{2}$ distribution so a Monte Carlo procedure is
employed.

Since the goodness of fit of the model cannot be assessed through the
usual statistics and Monte Car\-lo procedures are computationally
expensive, some sta\-tistics based on lower dimensional marginals of the
contingency table have been proposed. In general the statistics
proposed are quadratic forms of the residuals
%
%
\begin{equation}\label{eqgof}
\{\mathbf{p}_{r}-\bolds{\pi}_{r}(\hat
{\bolds{\psi}}) \}^{\prime} \mathbf{C} \{\mathbf{p}_{r}-\bolds{\pi
}_{r}(\hat{\bolds{\psi}}) \},
\end{equation}
where $\mathbf{C}$ is a weight matrix, $\mathbf{p}_{r}$ denotes
the sample marginal proportions and $r$ denotes a set of lower order
marginals.

\citet{Maydeu01} considers the statistic $G$ as in (\ref{eqG})
employed for estimation, which corresponds to setting $\mathbf
{C}=\mathbf{W}$ in (\ref{eqgof}) and $r$ denoting univariate and
bivariate marginal probabilities. The statistic $S \hat G$ is analyzed
in order to test $H_{0}\dvtx \bolds{\kappa}=\bolds{\kappa}(\bolds{\psi})$.
When $\hat{\mathbf{W}}=\hat{\bolds{\Xi}}^{-1}$, then $S \hat G \stackrel
{d}{\rightarrow} \chi^{2}_{d}$
where $d=N (N+1)/2-q$ and $q$ is the number of model parameters.
However, when $\hat{\mathbf{W}}=[\operatorname{diag}(\hat{\bolds{\Xi
}})]^{-1}$ or \mbox{$\hat{\mathbf{W}}=I$}, the asymptotic distribution
of the statistic is a weight\-ed sum of $d$ chi-square random variables
with one degree of freedom. \citet{Maydeu01} proposes to rescale the
test statistic in order to match the asymptotic chi-square
distribution. The same procedure is followed in the proposal for
testing $H_{0}\dvtx\bolds{\pi}_{2}=\bolds{\pi}_{2}(\bolds{\psi})$, where
$\bolds{\pi}_{2}$ is the vector of all univariate and
bivariate marginal probabilities. \citeauthor{Maydeu06}\break (\citeyear{Maydeu06}) considers the
testing of further hypotheses but the issue of the asymptotic
distribution being a~weighted sum of chi-square distributions remains.

\citet{MaydeuJoe05} consider testing the hypothesis $H_{0}\dvtx
\tilde{\bolds{\pi}}=\tilde{\bolds{\pi}}(\bolds{\psi})$ in a
multidimensional contingency table, where $\tilde{\bolds{\pi}}$ is
the $2^{N}$-dimensional vector of joint probabilities. Again, the use
of mar\-ginal residuals up to order $r$ is considered. Let $\bolds{\pi}$
denote a vector which stacks all the marginal probabilities:
univariate, bivariate, trivariate and so on. There is a one-to-one
correspondence between $\tilde{\bolds{\pi}}$ and $\bolds{\pi}$ so that
for a particular matrix $\bolds{\Lambda}$ of 0's and
1's $\bolds{\pi}=\bolds{\Lambda}\tilde{\bolds{\pi}}$. If
only marginal probabilities up to order~$r$ are considered, then
$\bolds{\pi}_{r}=\bolds{\Lambda}_{r} \tilde{\bolds{\pi}}$
for a sub-matrix $\bolds{\Lambda}_{r}$ of $\bolds{\Lambda}$.
Let $\bolds{\Delta}=\partial\tilde{\bolds{\pi}}/\partial
\bolds{\psi}$ and $\bolds{\Gamma}=\mathbf{E}-\tilde{\bolds{\pi}} \tilde
{\bolds{\pi}}^{\prime}$, where $\mathbf{E}=
\operatorname{diag}(\tilde{\bolds{\pi}})$. \citet{MaydeuJoe05}
propose the
statistic
%
%
\begin{equation}\label{eqgofM}\quad
\hspace*{15pt}M_{r}=S \{\mathbf{p}_{r}-\bolds{\pi}_{r}(\hat{\bolds{\psi
}}) \}^{\prime} \mathbf{C}_{r}(\hat{\bolds{\psi}}) \{\mathbf
{p}_{r}-\bolds{\pi}_{r}(\hat{\bolds{\psi}}) \},
\end{equation}
where $\mathbf{C}_{r}(\bolds{\psi})=\mathbf{F}_{r}^{-1}-\mathbf
{F}_{r}^{-1}\bolds{\Delta}_{r}(\bolds{\Delta}_{r}^{\prime}
\mathbf{F}_{r}^{-1}\bolds{\Delta}_{r})^{-1}\bolds{\Delta}_{r}^{\prime
}\mathbf{F}_{r}^{-1}$,
$\mathbf{F}_{r}=\bolds{\Lambda}_{r} \bolds{\Gamma}
\bolds{\Lambda}_{r}^{\prime}$ and $\bolds{\Delta}_{r}=\bolds{\Lambda
}_{r} \bolds{\Delta}$. $M_{r}$ is\vadjust{\goodbreak}
asymptotically distributed as a $\chi^{2}_{l-q}$ random\vspace*{1pt} variable where
$l$ is the length of $\mathbf{p}_{r}$. The $M_{r}$ statistic
asymptotically follows a chi-square distribution not only when
$\hat{\bolds{\psi}}$ is the maximum likelihood estimator, but also
when it is a~$\sqrt{S}$-consistent estimate, such as the limited
information estimator and the pairwise likelihood estimator presented
in Section \ref{secestimation}. Since the marginals should not be
sparse, \citet{MaydeuJoe05} suggest to use $M_{2}$ when the model is
identified using only univariate and bivariate information, also
because only up to bivariate sample moments and four-way model
probabilities are involved in the computation of $M_{2}$. As the number
of cells gets larger, the dimension of the matrices involved in
(\ref{eqgofM}) increases noticeably, and tricks may be necessary to do
the computations. Analysis and extensions of this type of test are
considered in \citet{MaydeuJoe06}, \citet{Reiser08} and
\citet{JoeMaydeu10}. All applications considered regard item response
theory, so an investigation of their performance in paired comparison
data is necessary to understand the sample size needed for obtaining
accurate Type I errors using $M_{2}$.

\section{Software}\label{sec6}

Fitting models to paired comparison data is facilitated by some
\texttt{R} packages which allow fitting of the classical models and, in
some cases, also fitting of more complicated models.

The \texttt{eba} package (\cite{Wickelmaier04}) fits elimination by
aspects models (\cite{Tversky72}) to paired comparison data. The
elimination by aspects model assumes that different objects present
various aspects. The worth of each object is the sum of the worth
associated with each aspect possessed by the object. When all objects
possess only one relevant aspect, then the elimination by aspects model
reduces to the Bradley--Terry model. Therefore, in case only one aspect
per object is specified, the function \texttt{eba} can be used to fit
model (\ref{LinMod}) with logit link, while when the link is probit the
function \texttt{thurstone} can be used. The function \texttt{strans}
checks how many violations of weak, moderate and strong stochastic
transitivity are present in the data.

The \texttt{prefmod} package (\cite{Hatzinger10}) fits Brad\-ley--Terry
models exploiting their log-linear representation. Ordinal paired
comparisons are allowed, but the software reduces the total number of
categories to three or two, depending on whether there is a no
preference category or not.

There are three different functions for estimating models for
paired\vadjust{\goodbreak}
comparison data: the \texttt{llbt.fit} function which estimates the
log-linear version of the Bradley--Terry model through the estimation
algorithm described in \citet{Hatzinger04}, the \texttt{llbtPC.fit}
function that estimates the log-linear model exploiting the
\texttt{gnm} (\cite{Turner10gnm}) function for fitting generalized
nonlinear models and the \texttt{pattPC.fit} function, which fits
paired comparison data using a pattern design, that is, all possible
patterns of paired comparisons. The latter function handles also some
cases in which the responses are missing not at random; see Section
\ref{sec7}. A difficulty of this approach is that the response table
grows dramatically with the number of objects since, in case of only
two possible outcomes, the number of patterns is $2^{N}$, so no more
than six objects can be included with two response categories, and not
more than five with three response categories. Finally, the function
\mbox{\texttt{pattnpml.fit}} fits a mixture model to overdispersed paired
comparison data using nonparametric maximum likelihood.

The \texttt{BradleyTerry2} package (\cite{Turner10}) expands the
previous \texttt{BradleyTerry} (\cite{Firth08}) package and allows one
to fit the unstructured model (\ref{LinMod}) and extension (\ref{cov})
with logit, probit and cauchit link functions, including also
com\-parison-specific covariates. Model fitting is either by maximum
likelihood, penalized quasi-likelihood or bias-reduced maximum
likelihood (\cite{Firth93}).
In case of object specific random effects, as in model (\ref{eqre}),
penalized quasi-likelihood (\cite{Breslow93}) is used, while when an
object wins or loses all the paired comparisons in which it is involved
and its estimate worth parameter is infinite, then the bias-reduced
maximum likelihood produces finite estimates. If there are missing
explanatory variables, an additional worth parameter for the object
with missing covariates is estimated. Order effects and more general
comparison-specific covariates can be included, but only win-loss
responses are allowed.

The package \texttt{psychotree} (\cite{Strobl10}) implements the method
for recursive partitioning of the subjects on the basis of their
explanatory variables and estimates an unstructured Bradley--Terry
model for each of the final subgroups of subjects; see Section
\ref{sec4}.

Although the available packages have many useful features, a
combination of those provided by the different packages and also some
additional features could be of practical help. The \texttt{prefmod}
and\break \texttt{BradleyTerry2} packages were built with the aim of
analyzing multiple judgment data and tournament-like\vadjust{\goodbreak}
data, respectively. This is reflected in the different characteristics
of the packages. A function that can handle data with at least
three-categorical results, thus allowing for the ``no preference''
category, include different link functions, and an easy implementation
of object-, subject- and comparison-specific covariates in a linear
model framework would be useful. The available methods for including
dependencies between observations are only in a log-linear framework
through the introduction of further parameters in the predictor or
including object-related random effects, which are estimated by means
of penalized quasi likelihood, a~method that does not perform well with
binary data. At present, there are no available packages for the
analysis of paired comparison data that allow the fitting of models as
those presented in Section \ref{secthurstone}. However, implementation
of pairwise likelihood estimation for those models is straightforward
since it implies only the computation of bivariate normal
probabilities.\vspace*{1pt}

\section{Conclusions}\label{sec7}\vspace*{1pt}

This paper reviews some of the extensions proposed in the
literature to the two most commonly applied models for paired
comparison data, namely the Bradley--Terry and the Thurstone models.
However, not every aspect could be considered here, and among issues
that have not been treated, there are the development of models for
multi-dimensional da\-ta when objects are evaluated with respect to
multiple aspects (\cite{Bockenholt88}; \cite{Dittrich06}), the temporal
extension for comparisons repeated in time (\cite{Fahrmeir94}, \cite{Glickman01}, \cite{Bockenholt02}, \cite{Dittrich03}),
the estimation of abilities
of individuals belonging to a team that performs the paired comparisons
(\cite{Huang06}; \cite{Menke08}) and many more. Another important issue
concerns the optimal design of the experiment. \citet{Grasshoff04}
show that the minimum sample size required for maximizing the
determinant of the information matrix in an unstructured Bradley--Terry
model requires that every comparison is performed once. When objects
are specified using factors with a certain number of levels, the
required sample size grows exponentially, while the number of
parameters grows linearly as the number of factors increases. Some
designs, in order to reduce the number of required comparisons, are
investigated in \citet{Grasshoff04}. In \citet{Grashoff08} a
characterization of the locally optimal design in case of two factors
design in a Bradley--Terry model is given, but for more complex
situations it seems difficult to give general results. \citet{Goos11}
consider also the problem when within-pair order effects are present.
It seems that investigation of these issues in other models are not
present in the literature.

The methods for independent data are well established, and a lot of
literature has been published about them. The problem of the asymptotic
behavior of the maximum likelihood estimator has been tackled. The case
of a fixed number of objects and increasing number of comparisons per
couple does not seem to pose particular difficulties for standard
arguments, while more problematic appears the instance of a fixed
number of comparisons per couple and increasing number of items. In the
context of the unstructured Bradley--Terry model, \citet{Simons99}
find a condition on the growth rate of the largest ratio between item
worth parameters which assures that the maximum likelihood estimator is
consistent and asymptotically normally distributed. \citet{Yan11}
investigate the case in which the number of comparisons per couple is
not fixed, and some comparisons may also be missing, and find a
condition that assures normality of the maximum likelihood estimator.
We are not acquainted with any other investigation of asymptotic
behavior of estimators in models different from the unstructured,
independent Bradley--Terry model.

Particular attention has been focused on models for dependent data.
Thurstonian models appear particularly suitable to account for
dependence between observations. However, the problems posed by the
identification restrictions are noticeable. The estimated model has to
be interpreted with reference to a class of covariance matrices, and
different identification restrictions may lead to different class of
matrices. It is possible to rotate the matrix according to a predefined
hypothesis about the covariance between certain items
(\cite{Maydeu07}), but the estimated standard errors vary depending on
the fixed parameters and the significance of the other estimated
parameters changes.

In the multiple judgment sampling scheme it is often stated that if a
judge does not perform all paired comparisons, then it suffices to
define subject-specific matrices $\mathbf{A}_{s}$ (see Section
\ref{secthurstone}) with rows corresponding only to the comparisons
performed by\break judge~$s$. However, it is expected that this may be
problematic for estimation by means of limited information
estimation,\vadjust{\goodbreak}
and there are no studies about the consequences of missing data in this
estimation method.

Missing observations cause problems also for testing the goodness of
fit since quadratic statistics as (\ref{eqgof}) assume that all
comparisons are performed by all subjects.

Missing data may derive from the design of the experiment, for example
when $n$ is very large, and only a subset of all comparisons is
presented to each subject. Otherwise, if many comparisons are performed
by the same subject it may be necessary to account for the fatigue of
subjects and/or for the passing of time when comparisons take long in
order to be accomplished.

\citet{Dittrich11} consider the problem of missing data in the context
of the log-linear representation of the Bradley--Terry model since the
study of the missing mechanism may shed light on the psychological
process. It is assumed that the probability that a comparison is
missing follows a logistic distribution since this facilitates the
fitting of the model. However, the likelihood for such models is not
easy to compute, and the function in the \texttt{prefmod} package
allows one to compute it only for data with up to six objects. It is
not easy to discriminate between different types of missing mechanisms,
and a very large number of observations may be needed in order to
discriminate between a missing completely at random and missing not at
random situation.

The economic theory points out some problems in choice data that have
not been considered yet. The main aspects which may need to be
incorporated in models include the influence that subjects can have on
each other, the influence of one particular subject, that may be some
sort of leader, over all the other judges and the dependence on choices
caused by the social and cultural context. Inclusions of these aspects
will inevitably lead to even more complicated models for paired
comparison data.

Finally, methods for object-related dependencies present many open
problems. Most of the issues are connected to the dependence among all
comparisons which is typically present in this context. Moreover, the
scheme of paired comparisons is often much less balanced than in
psychometric experiments. Asymptotic theory in models for dependent
data when the number of items compared increases has not been developed
yet. Maximum pairwise likelihood estimation provided encouraging
results, but more extensive studies seem necessary. In this case,
computation of standard errors is problematic since there are
no\vadjust{\goodbreak}
independent replications of the data, so a viable alternative lies in
parametric bootstrap. Methods for model selection and goodness of fit
described in Section \ref{secmodel} require independent replication of
all comparisons; hence they cannot be employed in this setting.\vspace*{1pt}

\section*{Acknowledgments}\vspace*{1pt}
The author would like to thank Cristiano Varin and Alessandra Salvan
for helpful comments and the anonymous referees and an Associate Editor
for comments and suggestions that led to a substantial improvement of
the manuscript.\vspace*{1pt}

%

%

\begin{thebibliography}{122}

\bibitem[\protect\citeauthoryear{Agresti}{1992}]{Agresti92}
%
\begin{barticle}[auto:STB|2012/06/04|06:16:18]
\bauthor{\bsnm{Agresti},~\bfnm{A.}\binits{A.}}
(\byear{1992}).
\btitle{Analysis of ordinal paired comparison data}.
\bjournal{J. R. Stat. Soc. Ser. C Appl. Stat.}
\bvolume{41}
\bpages{287--297}.
\bptok{imsref}%
\end{barticle}
%
\endbibitem

\bibitem[\protect\citeauthoryear{Agresti}{2002}]{Agresti02}
%
\begin{bbook}[mr]
\bauthor{\bsnm{Agresti},~\bfnm{Alan}\binits{A.}}
(\byear{2002}).
\btitle{Categorical Data Analysis},
\bedition{2nd} ed.
\bpublisher{Wiley}, \baddress{New York}.
\bid{doi={10.1002/0471249688}, mr={1914507}}
\bptok{imsref}%
\end{bbook}
%
\endbibitem

\bibitem[\protect\citeauthoryear{Barry and Hartigan}{1993}]{Barry93}
%
\begin{barticle}[mr]
\bauthor{\bsnm{Barry},~\bfnm{Daniel}\binits{D.}} \AND
\bauthor{\bsnm{Hartigan},~\bfnm{J.~A.}\binits{J.~A.}}
(\byear{1993}).
\btitle{Choice models for predicting divisional winners in major league baseball}.
\bjournal{J.~Amer. Statist. Assoc.}
\bvolume{88}
\bpages{766--774}.
\bid{issn={0162-1459}}
\bptok{imsref}%
\end{barticle}
%
\endbibitem

\bibitem[\protect\citeauthoryear{B{\"{a}}uml}{1994}]{Bauml94}
%
\begin{barticle}[pbm]
\bauthor{\bsnm{B{\"{a}}uml},~\bfnm{K.~H.}\binits{K.~H.}}
(\byear{1994}).
\btitle{Upright versus upside-down faces: How interface attractiveness varies
with orientation}.
\bjournal{Percept. Psychophys}.
\bvolume{56}
\bpages{163--172}.
\bid{issn={0031-5117}, pmid={7971117}}
\bptok{imsref}%
\end{barticle}
%
\endbibitem

\bibitem[\protect\citeauthoryear{B{\"o}ckenholt}{1988}]{Bockenholt88}
%
\begin{barticle}[mr]
\bauthor{\bsnm{B{\"o}ckenholt},~\bfnm{U.}\binits{U.}}
(\byear{1988}).
\btitle{A logistic representation of multivariate paired-comparison models}.
\bjournal{J. Math. Psych.}
\bvolume{32}
\bpages{44--63}.
\bid{doi={10.1016/0022-2496(88)90037-5}, issn={0022-2496}, mr={0935673}}
\bptok{imsref}%
\end{barticle}
%
\endbibitem

\bibitem[\protect\citeauthoryear{B{\"o}ckenholt}{2001a}]{Bockenholt01}
%
\begin{barticle}[auto:STB|2012/06/04|06:16:18]
\bauthor{\bsnm{B{\"o}ckenholt},~\bfnm{U.}\binits{U.}}
(\byear{2001}a).
\btitle{Hierarchical modeling of paired comparison data}.
\bjournal{Psychol. Methods}
\bvolume{6}
\bpages{49--66}.
\bptok{imsref}%
\end{barticle}
%
\endbibitem

\bibitem[\protect\citeauthoryear{B{\"o}ckenholt}{2001b}]{Bockenholt01b}
%
\begin{barticle}[auto:STB|2012/06/04|06:16:18]
\bauthor{\bsnm{B{\"o}ckenholt},~\bfnm{U.}\binits{U.}}
(\byear{2001}b).
\btitle{Thresholds and intransitivities in pairwise judgments: A multilevel
analysis}.
\bjournal{Journal of Educational and Behavioral Statistics}
\bvolume{26}
\bpages{269--282}.
\bptok{imsref}%
\end{barticle}
%
\endbibitem

\bibitem[\protect\citeauthoryear{B{\"o}ckenholt}{2002}]{Bockenholt02}
%
\begin{barticle}[mr]
\bauthor{\bsnm{B{\"o}ckenholt},~\bfnm{Ulf}\binits{U.}}
(\byear{2002}).
\btitle{A {T}hurstonian analysis of preference change}.
\bjournal{J. Math. Psych.}
\bvolume{46}
\bpages{300--314}.
\bid{doi={10.1006/jmps.2001.1389}, issn={0022-2496}, mr={1920807}}
\bptok{imsref}%
\end{barticle}
%
\endbibitem

\bibitem[\protect\citeauthoryear{B{\"{o}}ckenholt}{2004}]{Bockenholt04}
%
\begin{barticle}[pbm]
\bauthor{\bsnm{B{\"{o}}ckenholt},~\bfnm{Ulf}\binits{U.}}
(\byear{2004}).
\btitle{Comparative judgments as an alternative to ratings: Identifying the
scale origin}.
\bjournal{Psychol. Methods}
\bvolume{9}
\bpages{453--465}.
\bid{doi={10.1037/1082-989X.9.4.453}, issn={1082-989X}, pii={2004-21445-004},
pmid={15598099}}
\bptok{imsref}%
\end{barticle}
%
\endbibitem

\bibitem[\protect\citeauthoryear{B{\"o}ckenholt}{2006}]{Bockenholt06}
%
\begin{barticle}[mr]
\bauthor{\bsnm{B{\"o}ckenholt},~\bfnm{Ulf}\binits{U.}}
(\byear{2006}).
\btitle{Thurstonian-based analyses: Past, present, and future utilities}.
\bjournal{Psychometrika}
\bvolume{71}
\bpages{615--629}.
\bid{doi={10.1007/s11336-006-1598-5}, issn={0033-3123}, mr={2312235}}
\bptok{imsref}%
\end{barticle}
%
\endbibitem

\bibitem[\protect\citeauthoryear{B{\"o}ckenholt and Dillon}{1997a}]{Bockenholt97}
%
\begin{barticle}[auto:STB|2012/06/04|06:16:18]
\bauthor{\bsnm{B{\"o}ckenholt},~\bfnm{U.}\binits{U.}} \AND
\bauthor{\bsnm{Dillon},~\bfnm{W.~R.}\binits{W.~R.}}
(\byear{1997}a).
\btitle{Modeling within-subject dependencies in ordinal paired comparison
data}.
\bjournal{Psychometrika}
\bvolume{62}
\bpages{411--434}.
\bptok{imsref}%
\end{barticle}
%
\endbibitem

\bibitem[\protect\citeauthoryear{B{\"o}ckenholt and
Dillon}{1997b}]{BockenholtDillon97}
%
\begin{barticle}[auto:STB|2012/06/04|06:16:18]
\bauthor{\bsnm{B{\"o}ckenholt},~\bfnm{U.}\binits{U.}} \AND
\bauthor{\bsnm{Dillon},~\bfnm{W.~R.}\binits{W.~R.}}
(\byear{1997}b).
\btitle{Some new methods for an old problem: Modeling preference
changes and
competitive market structures in pretest market data}.
\bjournal{Journal of Marketing Research}
\bvolume{34}
\bpages{130--142}.
\bptok{imsref}%
\end{barticle}
%
\endbibitem

\bibitem[\protect\citeauthoryear{B{\"{o}}ckenholt and
Tsai}{2001}]{BockenholtTsai01}
%
\begin{barticle}[pbm]
\bauthor{\bsnm{B{\"{o}}ckenholt},~\bfnm{U.}\binits{U.}} \AND
\bauthor{\bsnm{Tsai},~\bfnm{R.~C.}\binits{R.~C.}}
(\byear{2001}).
\btitle{Individual differences in paired comparison data}.
\bjournal{Br. J. Math. Stat. Psychol.}
\bvolume{54}
\bpages{265--277}.
\bid{issn={0007-1102}, pmid={11817093}}
\bptok{imsref}%
\end{barticle}
%
\endbibitem

\bibitem[\protect\citeauthoryear{B{\"o}ckenholt and Tsai}{2007}]{Bockenholt07}
%
\begin{bincollection}[auto:STB|2012/06/04|06:16:18]
\bauthor{\bsnm{B{\"o}ckenholt},~\bfnm{U.}\binits{U.}} \AND
\bauthor{\bsnm{Tsai},~\bfnm{R.~C.}\binits{R.~C.}}
(\byear{2007}).
\btitle{Random-effects models for preference data}.
In \bbooktitle{Handbook of Statistics}
(\beditor{\bfnm{C.~R.}\binits{C.~R.}~\bsnm{Rao}} \AND
\beditor{\bfnm{S.}\binits{S.}~\bsnm{Sinharay}}, eds.)
\bvolume{26}
\bpages{447--468}.
\bpublisher{Elsevier}, \baddress{Amsterdam}.
\bptok{imsref}%
\end{bincollection}
%
\endbibitem

\bibitem[\protect\citeauthoryear{Bradley}{1976}]{Bradley76}
%
\begin{barticle}[mr]
\bauthor{\bsnm{Bradley},~\bfnm{Ralph~A.}\binits{R.~A.}}
(\byear{1976}).
\btitle{Science, statistics, and paired comparisons}.
\bjournal{Biometrics}
\bvolume{32}
\bpages{213--232}.
\bid{issn={0006-341X}, mr={0408132}}
\bptok{imsref}%
\end{barticle}
%
\endbibitem

\bibitem[\protect\citeauthoryear{Bradley and Terry}{1952}]{Bradley52}
%
\begin{barticle}[mr]
\bauthor{\bsnm{Bradley},~\bfnm{Ralph~Allan}\binits{R.~A.}} \AND
\bauthor{\bsnm{Terry},~\bfnm{Milton~E.}\binits{M.~E.}}
(\byear{1952}).
\btitle{Rank analysis of incomplete block designs. {I}. {T}he method of paired
comparisons}.
\bjournal{Biometrika}
\bvolume{39}
\bpages{324--345}.
\bid{issn={0006-3444}, mr={0070925}}
\bptok{imsref}%
\end{barticle}
%
\endbibitem

\bibitem[\protect\citeauthoryear{Breslow and Clayton}{1993}]{Breslow93}
%
\begin{barticle}[auto:STB|2012/06/04|06:16:18]
\bauthor{\bsnm{Breslow},~\bfnm{N.~E.}\binits{N.~E.}} \AND
\bauthor{\bsnm{Clayton},~\bfnm{D.~G.}\binits{D.~G.}}
(\byear{1993}).
\btitle{Approximate inference in generalized linear mixed models}.
\bjournal{J. Amer. Statist. Assoc.}
\bvolume{88}
\bpages{9--25}.
\bptok{imsref}%
\end{barticle}
%
\endbibitem

\bibitem[\protect\citeauthoryear{Carroll and De~Soete}{1991}]{Carroll91}
%
\begin{barticle}[auto:STB|2012/06/04|06:16:18]
\bauthor{\bsnm{Carroll},~\bfnm{J.~D.}\binits{J.~D.}} \AND
\bauthor{\bsnm{De~Soete},~\bfnm{G.}\binits{G.}}
(\byear{1991}).
\btitle{Toward a new paradigm for the study of multiattribute choice behavior.
Spatial and discrete modeling of pairwise preferences}.
\bjournal{American Psychologist}
\bvolume{46}
\bpages{342--351}.
\bptok{imsref}%
\end{barticle}
%
\endbibitem

\bibitem[\protect\citeauthoryear{Cattelan}{2009}]{Cattelan09}
%
\begin{bmisc}[auto:STB|2012/06/04|06:16:18]
\bauthor{\bsnm{Cattelan},~\bfnm{M.}\binits{M.}}
(\byear{2009}).
\bhowpublished{Correlation models for paired comparison data.
Ph.D. thesis, Dept. Statistical Sciences, Univ. Padua}.
\bptok{imsref}%
\end{bmisc}
%
\endbibitem

\bibitem[\protect\citeauthoryear{Cattelan, Varin and Firth}{2012}]{Cattelan10}
%
\begin{bmisc}[auto:STB|2012/06/04|06:16:18]
\bauthor{\bsnm{Cattelan},~\bfnm{M.}\binits{M.}},
\bauthor{\bsnm{Varin},~\bfnm{C.}\binits{C.}} \AND
\bauthor{\bsnm{Firth},~\bfnm{D.}\binits{D.}}
(\byear{2012}).
\bhowpublished{Dynamic Bradley--Terry modelling of sports tournaments.
\textit{J. R. Stat. Soc. Ser. C Appl. Stat.} To appear}.
\bptok{imsref}%
\end{bmisc}
%
\endbibitem

\bibitem[\protect\citeauthoryear{Causeur and Husson}{2005}]{Causeur05}
%
\begin{barticle}[mr]
\bauthor{\bsnm{Causeur},~\bfnm{David}\binits{D.}} \AND
\bauthor{\bsnm{Husson},~\bfnm{Fran{\c{c}}ois}\binits{F.}}
(\byear{2005}).
\btitle{A 2-dimensional extension of the {B}radley--{T}erry model for paired
comparisons}.
\bjournal{J.~Statist. Plann. Inference}
\bvolume{135}
\bpages{245--259}.
\bid{doi={10.1016/j.jspi.2004.05.008}, issn={0378-3758}, mr={2200468}}
\bptok{imsref}%
\end{barticle}
%
\endbibitem

\bibitem[\protect\citeauthoryear{Chib and Greenberg}{1998}]{Chib98}
%
\begin{barticle}[auto:STB|2012/06/04|06:16:18]
\bauthor{\bsnm{Chib},~\bfnm{S.}\binits{S.}} \AND
\bauthor{\bsnm{Greenberg},~\bfnm{E.}\binits{E.}}
(\byear{1998}).
\btitle{Analysis of multivariate probit models}.
\bjournal{Biometrika}
\bvolume{85}
\bpages{347--361}.
\bptok{imsref}%
\end{barticle}
%
\endbibitem

\bibitem[\protect\citeauthoryear{Choisel and Wickelmaier}{2007}]{Choisel07}
%
\begin{barticle}[pbm]
\bauthor{\bsnm{Choisel},~\bfnm{Sylvain}\binits{S.}} \AND
\bauthor{\bsnm{Wickelmaier},~\bfnm{Florian}\binits{F.}}
(\byear{2007}).
\btitle{Evaluation of multichannel reproduced sound: Scaling auditory
attributes underlying listener preference}.
\bjournal{J. Acoust. Soc. Am.}
\bvolume{121}
\bpages{388--400}.
\bid{issn={0001-4966}, pmid={17297794}}
\bptok{imsref}%
\end{barticle}
%
\endbibitem

\bibitem[\protect\citeauthoryear{Cox and Reid}{2004}]{Cox}
%
\begin{barticle}[mr]
\bauthor{\bsnm{Cox},~\bfnm{D.~R.}\binits{D.~R.}} \AND
\bauthor{\bsnm{Reid},~\bfnm{N.}\binits{N.}}
(\byear{2004}).
\btitle{A note on pseudolikelihood constructed from marginal densities}.
\bjournal{Biometrika}
\bvolume{91}
\bpages{729--737}.
\bid{doi={10.1093/biomet/91.3.729}, issn={0006-3444}, mr={2090633}}
\bptok{imsref}%
\end{barticle}
%
\endbibitem

\bibitem[\protect\citeauthoryear{Craig}{2008}]{Craig08}
%
\begin{barticle}[mr]
\bauthor{\bsnm{Craig},~\bfnm{Peter}\binits{P.}}
(\byear{2008}).
\btitle{A new reconstruction of multivariate normal orthant probabilities}.
\bjournal{J. R. Stat. Soc. Ser. B Stat. Methodol.}
\bvolume{70}
\bpages{227--243}.
\bid{doi={10.1111/j.1467-9868.2007.00625.x}, issn={1369-7412}, mr={2412640}}
\bptok{imsref}%
\end{barticle}
%
\endbibitem

\bibitem[\protect\citeauthoryear{David}{1988}]{David88}
%
\begin{bbook}[mr]
\bauthor{\bsnm{David},~\bfnm{H.~A.}\binits{H.~A.}}
(\byear{1988}).
\btitle{The Method of Paired Comparisons},
\bedition{2nd} ed.
\bseries{Griffin's Statistical Monographs \& Courses}
\bvolume{41}.
\bpublisher{Griffin}, \baddress{London}.
\bid{mr={0947340}}
\bptok{imsref}%
\end{bbook}
%
\endbibitem

\bibitem[\protect\citeauthoryear{Davidson}{1970}]{Davidson70}
%
\begin{barticle}[auto:STB|2012/06/04|06:16:18]
\bauthor{\bsnm{Davidson},~\bfnm{R.~R.}\binits{R.~R.}}
(\byear{1970}).
\btitle{On extending the Bradley--Terry model to accommodate ties in paired
comparison experiments}.
\bjournal{J. Amer. Statist. Assoc.}
\bvolume{65}
\bpages{317--328}.
\bptok{imsref}%
\end{barticle}
%
\endbibitem

\bibitem[\protect\citeauthoryear{Davidson and Farquhar}{1976}]{Davidson76}
%
\begin{barticle}[mr]
\bauthor{\bsnm{Davidson},~\bfnm{Roger~R.}\binits{R.~R.}} \AND
\bauthor{\bsnm{Farquhar},~\bfnm{Peter~H.}\binits{P.~H.}}
(\byear{1976}).
\btitle{A bibliography on the method of paired comparisons}.
\bjournal{Biometrics}
\bvolume{32}
\bpages{241--252}.
\bid{issn={0006-341X}, mr={0408134}}
\bptok{imsref}%
\end{barticle}
%
\endbibitem

\bibitem[\protect\citeauthoryear{De~Soete and Winsberg}{1993}]{DeSoete93}
%
\begin{barticle}[auto:STB|2012/06/04|06:16:18]
\bauthor{\bsnm{De~Soete},~\bfnm{G.}\binits{G.}} \AND
\bauthor{\bsnm{Winsberg},~\bfnm{S.}\binits{S.}}
(\byear{1993}).
\btitle{A Thurstonian pairwise choice model with univariate and multivariate
spline transformations}.
\bjournal{Psychometrika}
\bvolume{58}
\bpages{233--256}.
\bptok{imsref}%
\end{barticle}
%
\endbibitem

\bibitem[\protect\citeauthoryear{Dillon, Kumar and De~Borrero}{1993}]{Dillon93}
%
\begin{barticle}[auto:STB|2012/06/04|06:16:18]
\bauthor{\bsnm{Dillon},~\bfnm{W.~R.}\binits{W.~R.}},
\bauthor{\bsnm{Kumar},~\bfnm{A.}\binits{A.}} \AND
\bauthor{\bsnm{De~Borrero},~\bfnm{M.~S.}\binits{M.~S.}}
(\byear{1993}).
\btitle{Capturing individual differences in paired comparisons: An
extended BTL
model incorporating descriptor variables}.
\bjournal{Journal of Marketing Research}
\bvolume{30}
\bpages{42--51}.
\bptok{imsref}%
\end{barticle}
%
\endbibitem

\bibitem[\protect\citeauthoryear{Dittrich, Francis and
Katzenbeisser}{2008}]{Dittrich03}
%
\begin{bmisc}[auto:STB|2012/06/04|06:16:18]
\bauthor{\bsnm{Dittrich},~\bfnm{R.}\binits{R.}},
\bauthor{\bsnm{Francis},~\bfnm{B.}\binits{B.}} \AND
\bauthor{\bsnm{Katzenbeisser},~\bfnm{W.}\binits{W.}}
(\byear{2008}).
\bhowpublished{Temporal dependence in longitudinal paired comparisons.
Research report, Dept. Statistics and Mathematics, WU Vienna
Univ. Economics and Business}.
\bptok{imsref}%
\end{bmisc}
%
\endbibitem

\bibitem[\protect\citeauthoryear{Dittrich, Hatzinger and Katzenbeisser}{1998}]{Dittrich98}
%
\begin{barticle}[auto:STB|2012/06/04|06:16:18]
\bauthor{\bsnm{Dittrich},~\bfnm{R.}\binits{R.}},
\bauthor{\bsnm{Hatzinger},~\bfnm{R.}\binits{R.}} \AND
\bauthor{\bsnm{Katzenbeisser},~\bfnm{W.}\binits{W.}}
(\byear{1998}).
\btitle{Modelling the effect of subject-specific covariates in paired
comparison studies with an application to university rankings}.
\bjournal{J. R. Stat. Soc. Ser. C Appl. Stat.}
\bvolume{47}
\bpages{511--525}.
\bptok{imsref}%
\end{barticle}
%
\endbibitem

\bibitem[\protect\citeauthoryear{Dittrich, Hatzinger and
Katzenbeisser}{2001}]{Dittrich01}
%
\begin{barticle}[mr]
\bauthor{\bsnm{Dittrich},~\bfnm{R.}\binits{R.}},
\bauthor{\bsnm{Hatzinger},~\bfnm{R.}\binits{R.}} \AND
\bauthor{\bsnm{Katzenbeisser},~\bfnm{W.}\binits{W.}}
(\byear{2001}).
\btitle{Corrigendum: ``{M}odelling the effect of subject-specific
covariates in
paired comparison studies with an application to university rankings.''}
\bjournal{J. R. Stat. Soc. Ser. C Appl. Stat.}
\bvolume{50}
\bpages{247--249}.
\bid{doi={10.1111/1467-9876.00232}, issn={0035-9254}, mr={1833276}}
\bptok{imsref}%
\end{barticle}
%
\endbibitem

\bibitem[\protect\citeauthoryear{Dittrich, Hatzinger and
Katzenbeisser}{2002}]{Dittrich02}
%
\begin{barticle}[mr]
\bauthor{\bsnm{Dittrich},~\bfnm{R.}\binits{R.}},
\bauthor{\bsnm{Hatzinger},~\bfnm{R.}\binits{R.}} \AND
\bauthor{\bsnm{Katzenbeisser},~\bfnm{W.}\binits{W.}}
(\byear{2002}).
\btitle{Modelling dependencies in paired comparison data: A log-linear
approach}.
\bjournal{Comput. Statist. Data Anal.}
\bvolume{40}
\bpages{39--57}.
\bid{doi={10.1016/S0167-9473(01)00106-2}, issn={0167-9473}, mr={1921121}}
\bptok{imsref}%
\end{barticle}
%
\endbibitem

\bibitem[\protect\citeauthoryear{Dittrich, Hatzinger and
Katzenbeisser}{2004}]{Dittrich04}
%
\begin{barticle}[auto:STB|2012/06/04|06:16:18]
\bauthor{\bsnm{Dittrich},~\bfnm{R.}\binits{R.}},
\bauthor{\bsnm{Hatzinger},~\bfnm{R.}\binits{R.}} \AND
\bauthor{\bsnm{Katzenbeisser},~\bfnm{W.}\binits{W.}}
(\byear{2004}).
\btitle{A log-linear approach for modelling ordinal paired comparison
data on
motives to start a PhD program}.
\bjournal{Stat. Model.}
\bvolume{4}
\bpages{1--13}.
\bptok{imsref}%
\end{barticle}
%
\endbibitem

\bibitem[\protect\citeauthoryear{Dittrich et~al.}{2006}]{Dittrich06}
%
\begin{barticle}[mr]
\bauthor{\bsnm{Dittrich},~\bfnm{Regina}\binits{R.}},
\bauthor{\bsnm{Francis},~\bfnm{Brian}\binits{B.}},
\bauthor{\bsnm{Hatzinger},~\bfnm{Reinhold}\binits{R.}} \AND
\bauthor{\bsnm{Katzenbeisser},~\bfnm{Walter}\binits{W.}}
(\byear{2006}).
\btitle{Modelling dependency in multivariate paired comparisons: A log-linear
approach}.
\bjournal{Math. Social Sci.}
\bvolume{52}
\bpages{197--209}.
\bid{doi={10.1016/j.mathsocsci.2006.06.001}, issn={0165-4896}, mr={2257629}}
\bptok{imsref}%
\end{barticle}
%
\endbibitem

\bibitem[\protect\citeauthoryear{Dittrich et~al.}{2007}]{Dittrich07}
%
\begin{barticle}[mr]
\bauthor{\bsnm{Dittrich},~\bfnm{Regina}\binits{R.}},
\bauthor{\bsnm{Francis},~\bfnm{Brian}\binits{B.}},
\bauthor{\bsnm{Hatzinger},~\bfnm{Reinhold}\binits{R.}} \AND
\bauthor{\bsnm{Katzenbeisser},~\bfnm{Walter}\binits{W.}}
(\byear{2007}).
\btitle{A paired comparison approach for the analysis of sets of {L}ikert-scale
responses}.
\bjournal{Stat. Model.}
\bvolume{7}
\bpages{3--28}.
\bid{doi={10.1177/1471082X0600700102}, issn={1471-082X}, mr={2749821}}
\bptok{imsref}%
\end{barticle}
%
\endbibitem

\bibitem[\protect\citeauthoryear{Dittrich et~al.}{2012}]{Dittrich11}
%
\begin{barticle}[auto:STB|2012/06/04|06:16:18]
\bauthor{\bsnm{Dittrich},~\bfnm{R.}\binits{R.}},
\bauthor{\bsnm{Francis},~\bfnm{B.}\binits{B.}},
\bauthor{\bsnm{Hatzinger},~\bfnm{R.}\binits{R.}} \AND
\bauthor{\bsnm{Katzenbeisser},~\bfnm{W.}\binits{W.}}
(\byear{2012}).
\btitle{Missing observations in paired comparison data}.
\bjournal{Stat. Model.}
\bvolume{12}
\bpages{117--143}.
\bptok{imsref}%
\end{barticle}
%
\endbibitem

\bibitem[\protect\citeauthoryear{Duineveld, Arents and King}{2000}]{Duineveld00}
%
\begin{barticle}[auto:STB|2012/06/04|06:16:18]
\bauthor{\bsnm{Duineveld},~\bfnm{C.~A.~A.}\binits{C.~A.~A.}},
\bauthor{\bsnm{Arents},~\bfnm{P.}\binits{P.}} \AND
\bauthor{\bsnm{King},~\bfnm{B.~M.}\binits{B.~M.}}
(\byear{2000}).
\btitle{Log-linear modelling of paired comparison
data from consumer tests}.
\bjournal{Food Quality and Preference} \bvolume{11}
\bpages{63--70}.
\bptok{imsref}%
\end{barticle}
%
\endbibitem

\bibitem[\protect\citeauthoryear{Ellermeier, Mader and
Daniel}{2004}]{Ellermeier04}
%
\begin{barticle}[auto:STB|2012/06/04|06:16:18]
\bauthor{\bsnm{Ellermeier},~\bfnm{W.}\binits{W.}},
\bauthor{\bsnm{Mader},~\bfnm{M.}\binits{M.}} \AND
\bauthor{\bsnm{Daniel},~\bfnm{P.}\binits{P.}}
(\byear{2004}).
\btitle{Scaling the unpleasantness of sounds according to the BTL model:
Ratio-scale representation and psychoacoustical analysis}.
\bjournal{Acta Acustica United with Acustica}
\bvolume{90}
\bpages{101--\break107}.
\bptok{imsref}%
\end{barticle}
%
\endbibitem

\bibitem[\protect\citeauthoryear{Fahrmeir and Tutz}{1994}]{Fahrmeir94}
%
\begin{barticle}[auto:STB|2012/06/04|06:16:18]
\bauthor{\bsnm{Fahrmeir},~\bfnm{L.}\binits{L.}} \AND
\bauthor{\bsnm{Tutz},~\bfnm{G.}\binits{G.}}
(\byear{1994}).
\btitle{Dynamic stochastic models for time-dependent ordered paired comparison
systems}.
\bjournal{J. Amer. Statist. Assoc.}
\bvolume{89}
\bpages{1438--1449}.
\bptok{imsref}%
\end{barticle}
%
\endbibitem



\bibitem[\protect\citeauthoryear{Firth}{1993}]{Firth93}
%
\begin{barticle}[mr]
\bauthor{\bsnm{Firth},~\bfnm{David}\binits{D.}}
(\byear{1993}).
\btitle{Bias reduction of maximum likelihood estimates}.
\bjournal{Biometrika}
\bvolume{80}
\bpages{27--38}.
\bid{doi={10.1093/biomet/80.1.27}, issn={0006-3444}, mr={1225212}}
\bptok{imsref}%
\end{barticle}
%
\endbibitem

\bibitem[\protect\citeauthoryear{Firth}{2005}]{Firth05}
%
\begin{barticle}[auto:STB|2012/06/04|06:16:18]
\bauthor{\bsnm{Firth},~\bfnm{D.}\binits{D.}}
(\byear{2005}).
\btitle{Bradley--Terry models in {\tt R}}.
\bjournal{Journal of Statistical Software}
\bvolume{12}
\bpages{1--12}.
\bptok{imsref}%
\end{barticle}
%
\endbibitem

\bibitem[\protect\citeauthoryear{Firth}{2008}]{Firth08}
%
\begin{bmisc}[auto:STB|2012/06/04|06:16:18]
\bauthor{\bsnm{Firth},~\bfnm{D.}\binits{D.}}
(\byear{2008}).
\bhowpublished{\texttt{BradleyTerry}: Bradley--Terry models.
Available at
\texttt{\href{http://CRAN.R-project.org/package=BradleyTerry}{http://CRAN.R-project.org/package=}
\href{http://CRAN.R-project.org/package=BradleyTerry}{BradleyTerry}}}.
\bptok{imsref}%
\end{bmisc}
%
\endbibitem


\bibitem[\protect\citeauthoryear{Firth and de~Menezes}{2004}]{Firth04}
%
\begin{barticle}[mr]
\bauthor{\bsnm{Firth},~\bfnm{David}\binits{D.}} \AND\bauthor{\bparticle{de}
\bsnm{Menezes},~\bfnm{Ren{\'e}e~X.}\binits{R.~X.}}
(\byear{2004}).
\btitle{Quasi-variances}.
\bjournal{Biometrika}
\bvolume{91}
\bpages{65--80}.
\bid{doi={10.1093/biomet/91.1.65}, issn={0006-3444}, mr={2050460}}
\bptok{imsref}%
\end{barticle}
%
\endbibitem

\bibitem[\protect\citeauthoryear{Ford}{1957}]{Ford57}
%
\begin{barticle}[mr]
\bauthor{\bsnm{Ford},~\bfnm{L.~R.}\binits{L.~R.} \bsuffix{Jr.}}
(\byear{1957}).
\btitle{Solution of a ranking problem from binary comparisons}.
\bjournal{Amer. Math. Monthly}
\bvolume{64}
\bpages{28--33}.
\bid{issn={0002-9890}, mr={0097876}}
\bptok{imsref}%
\end{barticle}
%
\endbibitem

\bibitem[\protect\citeauthoryear{Francis, Dittrich and
Hatzinger}{2010}]{Francis10}
%
\begin{barticle}[mr]
\bauthor{\bsnm{Francis},~\bfnm{Brian}\binits{B.}},
\bauthor{\bsnm{Dittrich},~\bfnm{Regina}\binits{R.}} \AND
\bauthor{\bsnm{Hatzinger},~\bfnm{Reinhold}\binits{R.}}
(\byear{2010}).
\btitle{Modeling heterogeneity in ranked responses by nonparametric maximum
likelihood: How do {E}uropeans get their scientific knowledge?}
\bjournal{Ann. Appl. Stat.}
\bvolume{4}
\bpages{2181--2202}.
\bid{doi={10.1214/10-AOAS366}, issn={1932-6157}, mr={2829952}}
\bptok{imsref}%
\end{barticle}
%
\endbibitem

\bibitem[\protect\citeauthoryear{Francis et~al.}{2002}]{Francis02}
%
\begin{barticle}[mr]
\bauthor{\bsnm{Francis},~\bfnm{Brian}\binits{B.}},
\bauthor{\bsnm{Dittrich},~\bfnm{Regina}\binits{R.}},
\bauthor{\bsnm{Hatzinger},~\bfnm{Reinhold}\binits{R.}} \AND
\bauthor{\bsnm{Penn},~\bfnm{Roger}\binits{R.}}
(\byear{2002}).
\btitle{Analysing partial ranks by using smoothed paired comparison
methods: An
investigation of value orientation in {E}urope}.
\bjournal{J. R. Stat. Soc. Ser. C Appl. Stat.}
\bvolume{51}
\bpages{319--336}.
\bid{doi={10.1111/1467-9876.00271}, issn={0035-9254}, mr={1920800}}
\bptok{imsref}%
\end{barticle}
%
\endbibitem

\bibitem[\protect\citeauthoryear{Genz and Bretz}{2002}]{Genz02}
%
\begin{barticle}[mr]
\bauthor{\bsnm{Genz},~\bfnm{Alan}\binits{A.}} \AND
\bauthor{\bsnm{Bretz},~\bfnm{Frank}\binits{F.}}
(\byear{2002}).
\btitle{Comparison of methods for the computation of multivariate {$t$}
probabilities}.
\bjournal{J. Comput. Graph. Statist.}
\bvolume{11}
\bpages{950--971}.
\bid{doi={10.1198/106186002321018885}, issn={1061-8600}, mr={1944269}}
\bptok{imsref}%
\end{barticle}
%
\endbibitem

\bibitem[\protect\citeauthoryear{Glenn and David}{1960}]{Glenn60}
%
\begin{barticle}[auto:STB|2012/06/04|06:16:18]
\bauthor{\bsnm{Glenn},~\bfnm{W.~A.}\binits{W.~A.}} \AND
\bauthor{\bsnm{David},~\bfnm{H.~A.}\binits{H.~A.}}
(\byear{1960}).
\btitle{Ties in paired-comparison experiments using a modified
Thurstone--Mosteller model}.
\bjournal{Biometrics}
\bvolume{16}
\bpages{86--109}.
\bptok{imsref}%
\end{barticle}
%
\endbibitem

\bibitem[\protect\citeauthoryear{Glickman}{2001}]{Glickman01}
%
\begin{barticle}[mr]
\bauthor{\bsnm{Glickman},~\bfnm{Mark~E.}\binits{M.~E.}}
(\byear{2001}).
\btitle{Dynamic paired comparison models with stochastic variances}.
\bjournal{J. Appl. Stat.}
\bvolume{28}
\bpages{673--689}.
\bid{doi={10.1080/02664760120059219}, issn={0266-4763}, mr={1862491}}
\bptok{imsref}%
\end{barticle}
%
\endbibitem

\bibitem[\protect\citeauthoryear{Goos and Grossmann}{2011}]{Goos11}
%
\begin{barticle}[auto:STB|2012/06/04|06:16:18]
\bauthor{\bsnm{Goos},~\bfnm{P.}\binits{P.}} \AND
\bauthor{\bsnm{Grossmann},~\bfnm{H.}\binits{H.}}
(\byear{2011}).
\btitle{Optimal design of factorial paired comparison experiments in the
presence of within-pair order effects}.
\bjournal{Food Quality and Preference}
\bvolume{22}
\bpages{198--204}.
\bptok{imsref}%
\end{barticle}
%
\endbibitem

\bibitem[\protect\citeauthoryear{Gra{\normalfont\fontsize{11}{13}\selectfont{\ss}}hoff and Schwabe}{2008}]{Grashoff08}
%
\begin{barticle}[mr]
\bauthor{\bsnm{Gra{\normalfont\fontsize{7.2}{9.2}\selectfont{\ss}}hoff},~\bfnm{Ulrike}\binits{U.}} \AND
\bauthor{\bsnm{Schwabe},~\bfnm{Rainer}\binits{R.}}
(\byear{2008}).
\btitle{Optimal design for the {B}radley-{T}erry paired comparison model}.
\bjournal{Stat. Methods Appl.}
\bvolume{17}
\bpages{275--289}.
\bid{doi={10.1007/s10260-007-0058-4}, issn={1618-2510}, mr={2425186}}
\bptok{imsref}%
\end{barticle}
%
\endbibitem

\bibitem[\protect\citeauthoryear{Gra{\normalfont\fontsize{11}{13}\selectfont{\ss}}hoff et~al.}{2004}]{Grasshoff04}
%
\begin{barticle}[mr]
\bauthor{\bsnm{Gra{\normalfont\fontsize{7.2}{9.2}\selectfont{\ss}}hoff},~\bfnm{Ulrike}\binits{U.}},
\bauthor{\bsnm{Gro{\normalfont\fontsize{7.2}{9.2}\selectfont{\ss}}mann},~\bfnm{Heiko}\binits{H.}},
\bauthor{\bsnm{Holling},~\bfnm{Heinz}\binits{H.}} \AND
\bauthor{\bsnm{Schwabe},~\bfnm{Rainer}\binits{R.}}
(\byear{2004}).
\btitle{Optimal designs for main effects in linear paired comparison models}.
\bjournal{J. Statist. Plann. Inference}
\bvolume{126}
\bpages{361--376}.
\bid{doi={10.1016/j.jspi.2003.07.005}, issn={0378-3758}, mr={2090864}}
\bptok{imsref}%
\end{barticle}
%
\endbibitem

\bibitem[\protect\citeauthoryear{Hatzinger}{2010}]{Hatzinger10}
%
\begin{bmisc}[auto:STB|2012/06/04|06:16:18]
\bauthor{\bsnm{Hatzinger},~\bfnm{R.}\binits{R.}}
(\byear{2010}).
\bhowpublished{\texttt{prefmod}: Utilities to fit paired comparison models
for preferences. Available at
\texttt{\href{http://CRAN.R-project.org/package=prefmod}{http://CRAN.}
\href{http://CRAN.R-project.org/package=prefmod}{R-project.org/package=prefmod}}}.
\bptok{imsref}%
\end{bmisc}
%
\endbibitem

\bibitem[\protect\citeauthoryear{Hatzinger and Francis}{2004}]{Hatzinger04}
%
\begin{bmisc}[auto:STB|2012/06/04|06:16:18]
\bauthor{\bsnm{Hatzinger},~\bfnm{R.}\binits{R.}} \AND
\bauthor{\bsnm{Francis},~\bfnm{B.~J.}\binits{B.~J.}}
(\byear{2004}).
\bhowpublished{Fitting paired comparison models in R. Research report,
Univ. Wien. Available at
\url{http://epub.wu.ac.at/id/eprint/740}}.
\bptok{imsref}%
\end{bmisc}
%
\endbibitem

\bibitem[\protect\citeauthoryear{Head et~al.}{2008}]{Head08}
%
\begin{barticle}[auto:STB|2012/06/04|06:16:18]
\bauthor{\bsnm{Head},~\bfnm{M.~L.}\binits{M.~L.}},
\bauthor{\bsnm{Doughty},~\bfnm{P.}\binits{P.}},
\bauthor{\bsnm{Blomberg},~\bfnm{S.~P.}\binits{S.~P.}} \AND
\bauthor{\bsnm{Keogh},~\bfnm{S.}\binits{S.}}
(\byear{2008}).
\btitle{Chemical mediation of reciprocal mother--offspring recognition
in the
Southern Water Skink (\textit{Eulamprus heatwolei})}.
\bjournal{Australian Ecology}
\bvolume{33}
\bpages{20--28}.
\bptok{imsref}%
\end{barticle}
%
\endbibitem

\bibitem[\protect\citeauthoryear{Henery}{1992}]{Henery92}
%
\begin{barticle}[auto:STB|2012/06/04|06:16:18]
\bauthor{\bsnm{Henery},~\bfnm{R.~J.}\binits{R.~J.}}
(\byear{1992}).
\btitle{An extension to the Thurstone--Mosteller model for chess}.
\bjournal{The Statistician}
\bvolume{41}
\bpages{559--567}.
\bptok{imsref}%
\end{barticle}
%
\endbibitem

\bibitem[\protect\citeauthoryear{Huang, Weng and Lin}{2006}]{Huang06}
%
\begin{barticle}[mr]
\bauthor{\bsnm{Huang},~\bfnm{Tzu-Kuo}\binits{T.-K.}},
\bauthor{\bsnm{Weng},~\bfnm{Ruby~C.}\binits{R.~C.}} \AND
\bauthor{\bsnm{Lin},~\bfnm{Chih-Jen}\binits{C.-J.}}
(\byear{2006}).
\btitle{Generalized {B}radley-{T}erry models and multi-class probability
estimates}.
\bjournal{J.~Mach. Learn. Res.}
\bvolume{7}
\bpages{85--115}.
\bid{issn={1532-4435}, mr={2274363}}
\bptok{imsref}%
\end{barticle}
%
\endbibitem

\bibitem[\protect\citeauthoryear{Joe and Maydeu-Olivares}{2010}]{JoeMaydeu10}
%
\begin{barticle}[mr]
\bauthor{\bsnm{Joe},~\bfnm{Harry}\binits{H.}} \AND
\bauthor{\bsnm{Maydeu-Olivares},~\bfnm{Alberto}\binits{A.}}
(\byear{2010}).
\btitle{A general family of limited information goodness-of-fit
statistics for
multinomial data}.
\bjournal{Psychometrika}
\bvolume{75}
\bpages{393--419}.
\bid{doi={10.1007/s11336-010-9165-5}, issn={0033-3123}, mr={2719935}}
\bptok{imsref}%
\end{barticle}
%
\endbibitem

\bibitem[\protect\citeauthoryear{Kent}{1982}]{Kent82}
%
\begin{barticle}[mr]
\bauthor{\bsnm{Kent},~\bfnm{John~T.}\binits{J.~T.}}
(\byear{1982}).
\btitle{Robust properties of likelihood ratio tests}.
\bjournal{Biometrika}
\bvolume{69}
\bpages{19--27}.
\bid{doi={10.1093/biomet/69.1.19}, issn={0006-3444}, mr={0655667}}
\bptok{imsref}%
\end{barticle}
%
\endbibitem

\bibitem[\protect\citeauthoryear{Kissler and B{\"{a}}uml}{2000}]{Kissler00}
%
\begin{barticle}[pbm]
\bauthor{\bsnm{Kissler},~\bfnm{J.}\binits{J.}} \AND
\bauthor{\bsnm{B{\"{a}}uml},~\bfnm{K.~H.}\binits{K.~H.}}
(\byear{2000}).
\btitle{Effects of the beholder's age on the perception of facial
attractiveness}.
\bjournal{Acta Psychol. (Amst)}
\bvolume{104}
\bpages{145--166}.
\bid{issn={0001-6918}, pii={S0001-6918(00)00018-4}, pmid={10900703}}
\bptok{imsref}%
\end{barticle}
%
\endbibitem

\bibitem[\protect\citeauthoryear{Knorr-Held}{2000}]{Held00}
%
\begin{barticle}[auto:STB|2012/06/04|06:16:18]
\bauthor{\bsnm{Knorr-Held},~\bfnm{L.}\binits{L.}}
(\byear{2000}).
\btitle{Dynamic rating of sports teams}.
\bjournal{The Statistician}
\bvolume{49}
\bpages{261--276}.
\bptok{imsref}%
\end{barticle}
%
\endbibitem

\bibitem[\protect\citeauthoryear{Lancaster and Quade}{1983}]{Lancaster83}
%
\begin{barticle}[mr]
\bauthor{\bsnm{Lancaster},~\bfnm{J.~F.}\binits{J.~F.}} \AND
\bauthor{\bsnm{Quade},~\bfnm{Dana}\binits{D.}}
(\byear{1983}).
\btitle{Random effects in paired-comparison experiments using the
{B}radley--{T}erry model}.
\bjournal{Biometrics}
\bvolume{39}
\bpages{245--249}.
\bid{doi={10.2307/2530824}, issn={0006-341X}, mr={0712751}}
\bptok{imsref}%
\end{barticle}
%
\endbibitem

\bibitem[\protect\citeauthoryear{Le~Cessie and
Van~Houwelingen}{1994}]{LeCessie94}
%
\begin{barticle}[auto:STB|2012/06/04|06:16:18]
\bauthor{\bsnm{Le~Cessie},~\bfnm{S.}\binits{S.}} \AND
\bauthor{\bsnm{Van~Houwelingen},~\bfnm{J.~C.}\binits{J.~C.}}
(\byear{1994}).
\btitle{Logistic regression for correlated binary data}.
\bjournal{J. R. Stat. Soc. Ser. C Appl. Stat.}
\bvolume{43}
\bpages{95--108}.
\bptok{imsref}%
\end{barticle}
%
\endbibitem

\bibitem[\protect\citeauthoryear{Lele, Nadeem and Schmuland}{2010}]{Lele10}
%
\begin{barticle}[mr]
\bauthor{\bsnm{Lele},~\bfnm{Subhash~R.}\binits{S.~R.}},
\bauthor{\bsnm{Nadeem},~\bfnm{Khurram}\binits{K.}} \AND
\bauthor{\bsnm{Schmuland},~\bfnm{Byron}\binits{B.}}
(\byear{2010}).
\btitle{Estimability and likelihood inference for generalized linear mixed
models using data cloning}.
\bjournal{J. Amer. Statist. Assoc.}
\bvolume{105}
\bpages{1617--1625}.
\bid{doi={10.1198/jasa.2010.tm09757}, issn={0162-1459}, mr={2796576}}
\bptok{imsref}%
\end{barticle}
%
\endbibitem

\bibitem[\protect\citeauthoryear{Lindsay}{1988}]{Lindsay88}
%
\begin{bincollection}[mr]
\bauthor{\bsnm{Lindsay},~\bfnm{Bruce~G.}\binits{B.~G.}}
(\byear{1988}).
\btitle{Composite likelihood methods}.
In \bbooktitle{Statistical Inference from Stochastic Processes
({I}thaca, {NY},
1987)}.
\bseries{Contemp. Math.}
\bvolume{80}
\bpages{221--239}.
\bpublisher{Amer. Math. Soc.}, \baddress{Providence, RI}.
\bid{doi={10.1090/conm/080/999014}, mr={0999014}}
\bptok{imsref}%
\end{bincollection}
%
\endbibitem

\bibitem[\protect\citeauthoryear{Luce}{1959}]{Luce59}
%
\begin{bbook}[mr]
\bauthor{\bsnm{Luce},~\bfnm{R.~Duncan}\binits{R.~D.}}
(\byear{1959}).
\btitle{Individual Choice Behavior: {A} Theoretical Analysis}.
\bpublisher{Wiley}, \baddress{New York}.
\bid{mr={0108411}}
\bptok{imsref}%
\end{bbook}
%
\endbibitem

\bibitem[\protect\citeauthoryear{Marschak}{1960}]{Marschak60}
%
\begin{bincollection}[mr]
\bauthor{\bsnm{Marschak},~\bfnm{Jacob}\binits{J.}}
(\byear{1960}).
\btitle{Binary-choice constraints and random utility indicators}.
In \bbooktitle{Mathematical Methods in the Social Sciences, 1959}
(\beditor{\bsnm{Arrow},~\bfnm{K.~J.}\binits{K.~J.}},
\beditor{\bsnm{Karlin},~\bfnm{S.}\binits{S.}} \AND
\beditor{\bsnm{Suppes},~\bfnm{S.}\binits{S.}}, eds.)
\bpages{312--329}.
\bpublisher{Stanford Univ. Press}, \baddress{Stanford, CA}.
\bid{mr={0118556}}
\bptok{imsref}%
\end{bincollection}
%
\endbibitem

\bibitem[\protect\citeauthoryear{Matthews and Morris}{1995}]{Matthews95}
%
\begin{barticle}[auto:STB|2012/06/04|06:16:18]
\bauthor{\bsnm{Matthews},~\bfnm{J.~N.~S.}\binits{J.~N.~S.}} \AND
\bauthor{\bsnm{Morris},~\bfnm{K.~P.}\binits{K.~P.}}
(\byear{1995}).
\btitle{An application of Bradley--Terry-type models to the measurement of
pain}.
\bjournal{J. R. Stat. Soc. Ser. C Appl. Stat.}
\bvolume{44}
\bpages{243--255}.
\bptok{imsref}%
\end{barticle}
%
\endbibitem



\bibitem[\protect\citeauthoryear{Maydeu-Olivares}{2001}]{Maydeu01}
%
\begin{barticle}[mr]
\bauthor{\bsnm{Maydeu-Olivares},~\bfnm{Albert}\binits{A.}}
(\byear{2001}).
\btitle{Limited information estimation and testing of {T}hurstonian
models for
paired comparison data under multiple judgment sampling}.
\bjournal{Psychometrika}
\bvolume{66}
\bpages{209--227}.
\bid{doi={10.1007/BF02294836}, issn={0033-3123}, mr={1836935}}
\bptok{imsref}%
\end{barticle}
%
\endbibitem

\bibitem[\protect\citeauthoryear{Maydeu-Olivares}{2002}]{Maydeu02}
%
\begin{barticle}[mr]
\bauthor{\bsnm{Maydeu-Olivares},~\bfnm{Albert}\binits{A.}}
(\byear{2002}).
\btitle{Limited information estimation and testing of {T}hurstonian
models for
preference data}.
\bjournal{Math. Social Sci.}
\bvolume{43}
\bpages{467--483}.
\bid{doi={10.1016/S0165-4896(02)00017-3}, issn={0165-4896}, mr={2073576}}
\bptok{imsref}%
\end{barticle}
%
\endbibitem


\bibitem[\protect\citeauthoryear{Maydeu-Olivares}{2003}]{Maydeu03}
%
\begin{bmisc}[auto:STB|2012/06/04|06:16:18]
\bauthor{\bsnm{Maydeu-Olivares},~\bfnm{A.}\binits{A.}}
(\byear{2003}).
\bhowpublished{Thurstonian covariance and correlation structures for
multiple judgment paired comparison data. Working Papers Economia, Instituto
de Empresa, Area of Economic Environment. Available at
\url{http://econpapers.repec.org/RePEc:emp:wpaper:wp03-04}}.
\bptok{imsref}%
\end{bmisc}
%
\endbibitem


\bibitem[\protect\citeauthoryear{Maydeu-Olivares}{2006}]{Maydeu06}
%
\begin{barticle}[mr]
\bauthor{\bsnm{Maydeu-Olivares},~\bfnm{Albert}\binits{A.}}
(\byear{2006}).
\btitle{Limited information estimation and testing of discretized multivariate
normal structural models}.
\bjournal{Psychometrika}
\bvolume{71}
\bpages{57--77}.
\bid{doi={10.1007/s11336-005-0773-4}, issn={0033-3123}, mr={2272520}}
\bptok{imsref}%
\end{barticle}
%
\endbibitem

\bibitem[\protect\citeauthoryear{Maydeu-Olivares and B{\"o}ckenholt}{2005}]{Maydeu05}
%
\begin{barticle}[auto:STB|2012/06/04|06:16:18]
\bauthor{\bsnm{Maydeu-Olivares},~\bfnm{A.}\binits{A.}} \AND
\bauthor{\bsnm{B{\"o}ckenholt},~\bfnm{U.}\binits{U.}}
(\byear{2005}).
\btitle{Structural equation modeling of paired-comparison and ranking data}.
\bjournal{Psychometrika}
\bvolume{10}
\bpages{285--304}.
\bptok{imsref}%
\end{barticle}
%
\endbibitem

\bibitem[\protect\citeauthoryear{Maydeu-Olivares and
B{\"o}ckenholt}{2008}]{Maydeu08}
%
\begin{barticle}[auto:STB|2012/06/04|06:16:18]
\bauthor{\bsnm{Maydeu-Olivares},~\bfnm{A.}\binits{A.}} \AND
\bauthor{\bsnm{B{\"o}ckenholt},~\bfnm{U.}\binits{U.}}
(\byear{2008}).
\btitle{Modeling subject health outcomes. Top 10 reasons to use Thurstone's
method}.
\bjournal{Medical Care}
\bvolume{46}
\bpages{346--348}.
\bptok{imsref}%
\end{barticle}
%
\endbibitem

\bibitem[\protect\citeauthoryear{Maydeu-Olivares and Hern{\'a}ndez}{2007}]{Maydeu07}
%
\begin{barticle}[auto:STB|2012/06/04|06:16:18]
\bauthor{\bsnm{Maydeu-Olivares},~\bfnm{A.}\binits{A.}} \AND
\bauthor{\bsnm{Hern{\'a}ndez},~\bfnm{A.}\binits{A.}}
(\byear{2007}).
\btitle{Identification and small sample estimation of Thurstone's unrestricted
model for paired comparisons data}.
\bjournal{Multivariate Behavioral Research}
\bvolume{42}
\bpages{323--347}.
\bptok{imsref}%
\end{barticle}
%
\endbibitem

\bibitem[\protect\citeauthoryear{Maydeu-Olivares and Joe}{2005}]{MaydeuJoe05}
%
\begin{barticle}[mr]
\bauthor{\bsnm{Maydeu-Olivares},~\bfnm{Albert}\binits{A.}} \AND
\bauthor{\bsnm{Joe},~\bfnm{Harry}\binits{H.}}
(\byear{2005}).
\btitle{Limited- and full-information estimation and goodness-of-fit
testing in
{$2\sp n$} contingency tables: A unified framework}.
\bjournal{J. Amer. Statist. Assoc.}
\bvolume{100}
\bpages{1009--1020}.
\bid{doi={10.1198/016214504000002069}, issn={0162-1459}, mr={2201027}}
\bptok{imsref}%
\end{barticle}
%
\endbibitem

\bibitem[\protect\citeauthoryear{Maydeu-Olivares and Joe}{2006}]{MaydeuJoe06}
%
\begin{barticle}[mr]
\bauthor{\bsnm{Maydeu-Olivares},~\bfnm{Albert}\binits{A.}} \AND
\bauthor{\bsnm{Joe},~\bfnm{Harry}\binits{H.}}
(\byear{2006}).
\btitle{Limited information goodness-of-fit testing in multidimensional
contingency tables}.
\bjournal{Psychometrika}
\bvolume{71}
\bpages{713--732}.
\bid{doi={10.1007/s11336-005-1295-9}, issn={0033-3123}, mr={2312239}}
\bptok{imsref}%
\end{barticle}
%
\endbibitem

\bibitem[\protect\citeauthoryear{Mazzucchi, Linzey and Bruning}{2008}]{Mazzucchi08}
%
\begin{barticle}[auto:STB|2012/06/04|06:16:18]
\bauthor{\bsnm{Mazzucchi},~\bfnm{T.~A.}\binits{T.~A.}},
\bauthor{\bsnm{Linzey},~\bfnm{W.~G.}\binits{W.~G.}} \AND
\bauthor{\bsnm{Bruning},~\bfnm{A.}\binits{A.}}
(\byear{2008}).
\btitle{A paired comparison experiment for gathering expert judgment
for an
aircraft wiring risk assessment}.
\bjournal{Reliability Engineering and System Safety}
\bvolume{93}
\bpages{722--731}.
\bptok{imsref}%
\end{barticle}
%
\endbibitem

\bibitem[\protect\citeauthoryear{McFadden}{2001}]{McFadden01}
%
\begin{barticle}[auto:STB|2012/06/04|06:16:18]
\bauthor{\bsnm{McFadden},~\bfnm{D.}\binits{D.}}
(\byear{2001}).
\btitle{Economic choices}.
\bjournal{American Economic Review}
\bvolume{91}
\bpages{351--378}.
\bptok{imsref}%
\end{barticle}
%
\endbibitem

\bibitem[\protect\citeauthoryear{McHale and Morton}{2011}]{McHale11}
%
\begin{barticle}[auto:STB|2012/06/04|06:16:18]
\bauthor{\bsnm{McHale},~\bfnm{I.}\binits{I.}} \AND
\bauthor{\bsnm{Morton},~\bfnm{A.}\binits{A.}}
(\byear{2011}).
\btitle{A Bradley--Terry type model for forecasting tennis match results}.
\bjournal{International Journal of Forecasting}
\bvolume{27}
\bpages{619--630}.
\bptok{imsref}%
\end{barticle}
%
\endbibitem

\bibitem[\protect\citeauthoryear{Mease}{2003}]{Mease03}
%
\begin{barticle}[mr]
\bauthor{\bsnm{Mease},~\bfnm{David}\binits{D.}}
(\byear{2003}).
\btitle{A penalized maximum likelihood approach for the ranking of college
football teams independent of victory margins}.
\bjournal{Amer. Statist.}
\bvolume{57}
\bpages{241--248}.
\bid{doi={10.1198/0003130032396}, issn={0003-1305}, mr={2016258}}
\bptok{imsref}%
\end{barticle}
%
\endbibitem

\bibitem[\protect\citeauthoryear{Menke and Martinez}{2008}]{Menke08}
%
\begin{barticle}[auto:STB|2012/06/04|06:16:18]
\bauthor{\bsnm{Menke},~\bfnm{J.~E.}\binits{J.~E.}} \AND
\bauthor{\bsnm{Martinez},~\bfnm{T.~R.}\binits{T.~R.}}
(\byear{2008}).
\btitle{A Bradley--Terry artificial neural network model for individual ratings
in group competitions}.
\bjournal{Neural Computing \& Applications}
\bvolume{17}
\bpages{175--186}.
\bptok{imsref}%
\end{barticle}
%
\endbibitem

\bibitem[\protect\citeauthoryear{Miwa, Hayter and Kuriki}{2003}]{Miwa03}
%
\begin{barticle}[mr]
\bauthor{\bsnm{Miwa},~\bfnm{Tetsuhisa}\binits{T.}},
\bauthor{\bsnm{Hayter},~\bfnm{A.~J.}\binits{A.~J.}} \AND
\bauthor{\bsnm{Kuriki},~\bfnm{Satoshi}\binits{S.}}
(\byear{2003}).
\btitle{The evaluation of general non-centred orthant probabilities}.
\bjournal{J. R. Stat. Soc. Ser. B Stat. Methodol.}
\bvolume{65}
\bpages{223--234}.
\bid{doi={10.1111/1467-9868.00382}, issn={1369-7412}, mr={1959823}}
\bptok{imsref}%
\end{barticle}
%
\endbibitem

\bibitem[\protect\citeauthoryear{Molenberghs and Verbeke}{2005}]{Molenberghs05}
%
\begin{bbook}[mr]
\bauthor{\bsnm{Molenberghs},~\bfnm{Geert}\binits{G.}} \AND
\bauthor{\bsnm{Verbeke},~\bfnm{Geert}\binits{G.}}
(\byear{2005}).
\btitle{Models for Discrete Longitudinal Data}.
\bpublisher{Springer}, \baddress{New York}.
\bid{mr={2171048}}
\bptok{imsref}%
\end{bbook}
%
\endbibitem

\bibitem[\protect\citeauthoryear{Mosteller}{1951}]{Mosteller51}
%
\begin{barticle}[auto:STB|2012/06/04|06:16:18]
\bauthor{\bsnm{Mosteller},~\bfnm{F.}\binits{F.}}
(\byear{1951}).
\btitle{Remarks on the method of paired comparisons. I. The least squares
solution assuming equal standard deviations and equal correlations.
II. The
effect of an aberrant\vadjust{\goodbreak} standard deviation when equal standard
deviations and
equal correlations are assumed. III. A test of significance for paired
comparisons when equal standard deviations and equal correlations are
assumed}.
\bjournal{Psychometrika}
\bvolume{16}
\bpages{3--9, 203--218}.
\bptok{imsref}%
\end{barticle}
%
\endbibitem

\bibitem[\protect\citeauthoryear{Muth{\'e}n}{1978}]{Muthen78}
%
\begin{barticle}[mr]
\bauthor{\bsnm{Muth{\'e}n},~\bfnm{Bengt}\binits{B.}}
(\byear{1978}).
\btitle{Contributions to factor analysis of dichotomous variables}.
\bjournal{Psychometrika}
\bvolume{43}
\bpages{551--560}.
\bid{doi={10.1007/BF02293813}, issn={0033-3123}, mr={0521904}}
\bptok{imsref}%
\end{barticle}
%
\endbibitem

\bibitem[\protect\citeauthoryear{Muth{\'e}n}{1993}]{Muthen93}
%
\begin{bincollection}[auto:STB|2012/06/04|06:16:18]
\bauthor{\bsnm{Muth{\'e}n},~\bfnm{B.}\binits{B.}}
(\byear{1993}).
\btitle{Goodness of fit with categorical and other non normal variables}.
In \bbooktitle{Structural Equation Models}
(\beditor{\binits{K. A.} \bsnm{Bollen}},
\beditor{\binits{J. S.} \bsnm{Long}}, eds.).
\bpages{205--234}.
\bpublisher{Sage}, \baddress{Newbury Park, CA}.
\bptok{imsref}%
\end{bincollection}
%
\endbibitem

\bibitem[\protect\citeauthoryear{Muth{\'e}n, Du~Toit and
Spisic}{1997}]{Muthen97}
%
\begin{bmisc}[auto:STB|2012/06/04|06:16:18]
\bauthor{\bsnm{Muth{\'e}n},~\bfnm{B.}\binits{B.}},
\bauthor{\bsnm{Du~Toit},~\bfnm{S.~H.~C.}\binits{S.~H.~C.}} \AND
\bauthor{\bsnm{Spisic},~\bfnm{D.}\binits{D.}}
(\byear{1997}).
\bhowpublished{Robust inference using weighted least squares and
quadratic estimating equations in latent variable modeling with categorical
and continuous outcomes. Technical report}.
\bptok{imsref}%
\end{bmisc}
%
\endbibitem

\bibitem[\protect\citeauthoryear{R Development Core Team}{2011}]{R}
%
\begin{bmisc}[auto:STB|2012/06/04|06:16:18]
\borganization{R Development Core Team}
(\byear{2011}).
\bhowpublished{\texttt{R}: \textit{A Language and Environment for Statistical Computing}.
R Foundation for Statistical Computing, Vienna, Austria.
ISBN 3-900051-07-0. Available at \url{http://www.R-project.org}}.
\bptok{imsref}%
\end{bmisc}
%
\endbibitem

\bibitem[\protect\citeauthoryear{Rao and Kupper}{1967}]{Rao67}
%
\begin{barticle}[mr]
\bauthor{\bsnm{Rao},~\bfnm{P.~V.}\binits{P.~V.}} \AND
\bauthor{\bsnm{Kupper},~\bfnm{L.~L.}\binits{L.~L.}}
(\byear{1967}).
\btitle{Ties in paired-comparison experiments: {A} generalization of the
{B}radley--{T}erry model}.
\bjournal{J.~Amer. Statist. Assoc.}
\bvolume{62}
\bpages{194--204}.
\bid{issn={0162-1459}, mr={0217963}}
\bptok{imsref}%
\end{barticle}
%
\endbibitem

\bibitem[\protect\citeauthoryear{Reiser}{2008}]{Reiser08}
%
\begin{barticle}[mr]
\bauthor{\bsnm{Reiser},~\bfnm{Mark}\binits{M.}}
(\byear{2008}).
\btitle{Goodness-of-fit testing using components based on marginal frequencies
of multinomial data}.
\bjournal{British J. Math. Statist. Psych.}
\bvolume{61}
\bpages{331--360}.
\bid{doi={10.1348/000711007X204215}, issn={0007-1102}, mr={2649040}}
\bptok{imsref}%
\end{barticle}
%
\endbibitem

\bibitem[\protect\citeauthoryear{Sham and Curtis}{1995}]{Sham95}
%
\begin{barticle}[pbm]
\bauthor{\bsnm{Sham},~\bfnm{P.~C.}\binits{P.~C.}} \AND
\bauthor{\bsnm{Curtis},~\bfnm{D.}\binits{D.}}
(\byear{1995}).
\btitle{An extended transmission/disequilibrium test (TDT) for multi-allele
marker loci}.
\bjournal{Ann. Hum. Genet.}
\bvolume{59}
\bpages{323--336}.
\bid{issn={0003-4800}, pmid={7486838}}
\bptok{imsref}%
\end{barticle}
%
\endbibitem

\bibitem[\protect\citeauthoryear{Simons and Yao}{1999}]{Simons99}
%
\begin{barticle}[mr]
\bauthor{\bsnm{Simons},~\bfnm{Gordon}\binits{G.}} \AND
\bauthor{\bsnm{Yao},~\bfnm{Yi-Ching}\binits{Y.-C.}}
(\byear{1999}).
\btitle{Asymptotics when the number of parameters tends to infinity in the
{B}radley--{T}erry model for paired comparisons}.
\bjournal{Ann. Statist.}
\bvolume{27}
\bpages{1041--1060}.
\bid{doi={10.1214/aos/1018031267}, issn={0090-5364}, mr={1724040}}
\bptok{imsref}%
\end{barticle}
%
\endbibitem

\bibitem[\protect\citeauthoryear{Springall}{1973}]{Springall73}
%
\begin{barticle}[auto:STB|2012/06/04|06:16:18]
\bauthor{\bsnm{Springall},~\bfnm{A.}\binits{A.}}
(\byear{1973}).
\btitle{Response surface fitting using a generalization of the Bradley--Terry
paired comparison model}.
\bjournal{J. R. Stat. Soc. Ser. C Appl. Stat.}
\bvolume{22}
\bpages{59--68}.
\bptok{imsref}%
\end{barticle}
%
\endbibitem

\bibitem[\protect\citeauthoryear{Stern}{1990}]{Stern90}
%
\begin{barticle}[mr]
\bauthor{\bsnm{Stern},~\bfnm{Hal}\binits{H.}}
(\byear{1990}).
\btitle{A continuum of paired comparisons models}.
\bjournal{Biometrika}
\bvolume{77}
\bpages{265--273}.
\bid{doi={10.1093/biomet/77.2.265}, issn={0006-3444}, mr={1064798}}
\bptok{imsref}%
\end{barticle}
%
\endbibitem

\bibitem[\protect\citeauthoryear{Stern}{2011}]{Stern11}
%
\begin{barticle}[mr]
\bauthor{\bsnm{Stern},~\bfnm{Steven~E.}\binits{S.~E.}}
(\byear{2011}).
\btitle{Moderated paired comparisons: A generalized {B}radley--{T}erry
model for
continuous data using a discontinuous penalized likelihood function}.
\bjournal{J. R. Stat. Soc. Ser. C Appl. Stat.}
\bvolume{60}
\bpages{397--415}.
\bid{doi={10.1111/j.1467-9876.2010.00751.x}, issn={0035-9254}, mr={2767853}}
\bptok{imsref}%
\end{barticle}
%
\endbibitem

\bibitem[\protect\citeauthoryear{Stigler}{1994}]{Stigler94}
%
\begin{barticle}[auto:STB|2012/06/04|06:16:18]
\bauthor{\bsnm{Stigler},~\bfnm{S.~M.}\binits{S.~M.}}
(\byear{1994}).
\btitle{Citation patterns in the journals of statistics and probability}.
\bjournal{Statist. Sci.}
\bvolume{9}
\bpages{94--108}.
\bptok{imsref}%
\end{barticle}
%
\endbibitem

\bibitem[\protect\citeauthoryear{Strobl, Wickelmaier and Zeileis}{2011}]{Strobl10}
%
\begin{barticle}[auto:STB|2012/06/04|06:16:18]
\bauthor{\bsnm{Strobl},~\bfnm{C.}\binits{C.}},
\bauthor{\bsnm{Wickelmaier},~\bfnm{F.}\binits{F.}} \AND
\bauthor{\bsnm{Zeileis},~\bfnm{A.}\binits{A.}}
(\byear{2011}).
\btitle{Accounting for individual differences in Bradley--Terry models
by means
of recursive partitioning}.
\bjournal{Journal of Educational and Behavioral Statistics}
\bvolume{36}
\bpages{135--153}.
\bptok{imsref}%
\end{barticle}
%
\endbibitem

\bibitem[\protect\citeauthoryear{Stuart-Fox et~al.}{2006}]{Stuart06}
%
\begin{barticle}[auto:STB|2012/06/04|06:16:18]
\bauthor{\bsnm{Stuart-Fox},~\bfnm{D.~M.}\binits{D.~M.}},
\bauthor{\bsnm{Firth},~\bfnm{D.}\binits{D.}},
\bauthor{\bsnm{Moussalli},~\bfnm{A.}\binits{A.}} \AND
\bauthor{\bsnm{\mbox{Whiting}},~\bfnm{M.~J.}\binits{M.~J.}}
(\byear{2006}).
\btitle{Multiple signals in chameleon contests: Designing and analysing animal
contests as a tournament}.
\bjournal{Animal Behavior}
\bvolume{71}
\bpages{1263--1271}.
\bptok{imsref}%
\end{barticle}
%
\endbibitem

\bibitem[\protect\citeauthoryear{Takane}{1989}]{Takane89}
%
\begin{bincollection}[auto:STB|2012/06/04|06:16:18]
\bauthor{\bsnm{Takane},~\bfnm{Y.}\binits{Y.}}
(\byear{1989}).
\btitle{Analysis of covariance structures and probabilistic binary choice
data}.
In \bbooktitle{New Developments in Psychological Choice Modeling}
(\beditor{\bfnm{G.}\binits{G.}~\bsnm{De~Soete}},
\beditor{\bfnm{H.}\binits{H.}~\bsnm{Feger}} \AND
\beditor{\bfnm{K.~C.}\binits{K.~C.}~\bsnm{Klauser}}, eds.).
\bpublisher{North-Holland}, \baddress{Amsterdam}.
\bptok{imsref}%
\end{bincollection}
%
\endbibitem

\bibitem[\protect\citeauthoryear{Thurstone}{1927}]{Thurstone27}
%
\begin{barticle}[auto:STB|2012/06/04|06:16:18]
\bauthor{\bsnm{Thurstone},~\bfnm{L.~L.}\binits{L.~L.}}
(\byear{1927}).
\btitle{A law of comparative judgment}.
\bjournal{Psychological Review}
\bvolume{34}
\bpages{368--389}.
\bptok{imsref}%
\end{barticle}
%
\endbibitem

\bibitem[\protect\citeauthoryear{Thurstone and Jones}{1957}]{Thurstone57}
%
\begin{barticle}[auto:STB|2012/06/04|06:16:18]
\bauthor{\bsnm{Thurstone},~\bfnm{L.~L.}\binits{L.~L.}} \AND
\bauthor{\bsnm{Jones},~\bfnm{L.~V.}\binits{L.~V.}}
(\byear{1957}).
\btitle{The rational origin for measuring subjective values}.
\bjournal{J. Amer. Statist. Assoc.}
\bvolume{52}
\bpages{458--471}.
\bptok{imsref}%
\end{barticle}
%
\endbibitem

\bibitem[\protect\citeauthoryear{Train}{2009}]{Train09}
%
\begin{bbook}[mr]
\bauthor{\bsnm{Train},~\bfnm{Kenneth~E.}\binits{K.~E.}}
(\byear{2009}).
\btitle{Discrete Choice Methods with Simulation}, \bedition{2nd} ed.
\bpublisher{Cambridge Univ. Press}, \baddress{Cambridge}.
\bid{mr={2519514}}
\bptok{imsref}%
\end{bbook}
%
\endbibitem

\bibitem[\protect\citeauthoryear{Tsai}{2000}]{Tsai00}
%
\begin{barticle}[mr]
\bauthor{\bsnm{Tsai},~\bfnm{Rung-Ching}\binits{R.-C.}}
(\byear{2000}).
\btitle{Remarks on the identifiability of {T}hurstonian ranking models: {C}ase
{V}, {C}ase {III}, or neither?}
\bjournal{Psychometrika}
\bvolume{65}
\bpages{233--240}.
\bid{doi={10.1007/BF02294376}, issn={0033-3123}, mr={1763521}}
\bptok{imsref}%
\end{barticle}
%
\endbibitem

\bibitem[\protect\citeauthoryear{Tsai}{2003}]{Tsai03}
%
\begin{barticle}[mr]
\bauthor{\bsnm{Tsai},~\bfnm{Rung-Ching}\binits{R.-C.}}
(\byear{2003}).
\btitle{Remarks on the identifiability of {T}hurstonian paired comparison
models under multiple judgment}.
\bjournal{Psychometrika}
\bvolume{68}
\bpages{361--372}.
\bid{doi={10.1007/BF02294732}, issn={0033-3123}, mr={2272384}}
\bptok{imsref}%
\end{barticle}
%
\endbibitem

\bibitem[\protect\citeauthoryear{Tsai and B{\"o}ckenholt}{2002}]{Tsai02}
%
\begin{barticle}[mr]
\bauthor{\bsnm{Tsai},~\bfnm{Rung-Ching}\binits{R.-C.}} \AND
\bauthor{\bsnm{B{\"o}ckenholt},~\bfnm{Ulf}\binits{U.}}
(\byear{2002}).
\btitle{Two-level linear paired comparison models: Estimation and
identifiability issues}.
\bjournal{Math. Social Sci.}
\bvolume{43}
\bpages{429--449}.
\bid{doi={10.1016/S0165-4896(02)00019-7}, issn={0165-4896}, mr={2072966}}
\bptok{imsref}%
\end{barticle}
%
\endbibitem

\bibitem[\protect\citeauthoryear{Tsai and B{\"o}ckenholt}{2006}]{Tsai06}
%
\begin{barticle}[mr]
\bauthor{\bsnm{Tsai},~\bfnm{Rung-Ching}\binits{R.-C.}} \AND
\bauthor{\bsnm{B{\"o}ckenholt},~\bfnm{Ulf}\binits{U.}}
(\byear{2006}).
\btitle{Modelling intransitive preferences: A random-effects approach}.
\bjournal{J. Math. Psych.}
\bvolume{50}
\bpages{1--14}.
\bid{doi={10.1016/j.jmp.2005.11.004}, issn={0022-2496}, mr={2208061}}
\bptok{imsref}%
\end{barticle}
%
\endbibitem

\bibitem[\protect\citeauthoryear{Tsai and B{\"o}ckenholt}{2008}]{Tsai08}
%
\begin{barticle}[mr]
\bauthor{\bsnm{Tsai},~\bfnm{Rung-Ching}\binits{R.-C.}} \AND
\bauthor{\bsnm{B{\"o}ckenholt},~\bfnm{Ulf}\binits{U.}}
(\byear{2008}).
\btitle{On the importance of distinguishing between within- and between-subject
effects in intransitive intertemporal choice}.
\bjournal{J. Math. Psych.}
\bvolume{52}
\bpages{10--20}.
\bid{doi={10.1016/j.jmp.2007.09.004}, issn={0022-2496}, mr={2407792}}
\bptok{imsref}%
\end{barticle}
%
\endbibitem

\bibitem[\protect\citeauthoryear{Turner and Firth}{2010a}]{Turner10}
%
\begin{bmisc}[auto:STB|2012/06/04|06:16:18]
\bauthor{\bsnm{Turner},~\bfnm{H.}\binits{H.}} \AND
\bauthor{\bsnm{Firth},~\bfnm{D.}\binits{D.}}
(\byear{2010a}).
\bhowpublished{Bradley--Terry models in R: The \texttt{BradleyTerry2}
package. Available at
\texttt{\href{http://CRAN.R-project.org/package=BradleyTerry2}{http://}
\href{http://CRAN.R-project.org/package=BradleyTerry2}{CRAN.R-project.org/package=BradleyTerry2}}}.
\bptok{imsref}%
\end{bmisc}
%
\endbibitem\

\bibitem[\protect\citeauthoryear{Turner and Firth}{2010b}]{Turner10gnm}
%
\begin{bmisc}[auto:STB|2012/06/04|06:16:18]
\bauthor{\bsnm{Turner},~\bfnm{H.}\binits{H.}} \AND
\bauthor{\bsnm{Firth},~\bfnm{D.}\binits{D.}}
(\byear{2010b}).
\bhowpublished{Generalized nonlinear models in R: An overview of the
\texttt{gnm} package. Available at \href{http://CRAN.R-project.org/package=gnm}{http:// CRAN.R-project.org/package=gnm}}.
\bptok{imsref}%
\end{bmisc}
%
\endbibitem

\bibitem[\protect\citeauthoryear{Tversky}{1972}]{Tversky72}
%
\begin{barticle}[auto:STB|2012/06/04|06:16:18]
\bauthor{\bsnm{Tversky},~\bfnm{A.}\binits{A.}}
(\byear{1972}).
\btitle{Elimination by aspects: A theory of choice}.
\bjournal{Psychological Review}
\bvolume{79}
\bpages{281--299}.
\bptok{imsref}%
\end{barticle}
%
\endbibitem

\bibitem[\protect\citeauthoryear{Usami}{2010}]{Usami10}
%
\begin{barticle}[auto:STB|2012/06/04|06:16:18]
\bauthor{\bsnm{Usami},~\bfnm{S.}\binits{S.}}
(\byear{2010}).
\btitle{Individual differences multidimensional Bradley--Terry model using
reversible jump Markov chain Monte Carlo algorithm}.
\bjournal{Behaviormetrika}
\bvolume{37}
\bpages{135--\break155}.
\bptok{imsref}%
\end{barticle}
%
\endbibitem

\bibitem[\protect\citeauthoryear{Varin, Reid and Firth}{2011}]{Varin11}
%
\begin{barticle}[mr]
\bauthor{\bsnm{Varin},~\bfnm{Cristiano}\binits{C.}},
\bauthor{\bsnm{Reid},~\bfnm{Nancy}\binits{N.}} \AND
\bauthor{\bsnm{Firth},~\bfnm{David}\binits{D.}}
(\byear{2011}).
\btitle{An overview of composite likelihood methods}.
\bjournal{Statist. Sinica}
\bvolume{21}
\bpages{5--42}.
\bid{issn={1017-0405}, mr={2796852}}
\bptok{imsref}%
\end{barticle}
%
\endbibitem

\bibitem[\protect\citeauthoryear{Walker and Ben-Akiva}{2002}]{Walker02}
%
\begin{barticle}[mr]
\bauthor{\bsnm{Walker},~\bfnm{Joan}\binits{J.}} \AND
\bauthor{\bsnm{Ben-Akiva},~\bfnm{Moshe}\binits{M.}}
(\byear{2002}).
\btitle{Generalized random utility model}.
\bjournal{Math. Social Sci.}
\bvolume{43}
\bpages{303--343}.
\bid{doi={10.1016/S0165-4896(02)00023-9}, issn={0165-4896}, mr={2072961}}
\bptok{imsref}%
\end{barticle}
%
\endbibitem

\bibitem[\protect\citeauthoryear{Whiting et~al.}{2006}]{Whiting06}
%
\begin{barticle}[auto:STB|2012/06/04|06:16:18]
\bauthor{\bsnm{Whiting},~\bfnm{M.~J.}\binits{M.~J.}},
\bauthor{\bsnm{Stuart-Fox},~\bfnm{D.~M.}\binits{D.~M.}},
\bauthor{\bsnm{O'Connor},~\bfnm{D.}\binits{D.}},
\bauthor{\bsnm{Firth},~\bfnm{D.}\binits{D.}},
\bauthor{\bsnm{Bennett},~\bfnm{N.~C.}\binits{N.~C.}} \AND
\bauthor{\bsnm{Blomberg},~\bfnm{S.~P.}\binits{S.~P.}}
(\byear{2006}).
\btitle{Ultraviolet signals ultra-aggression in a lizard}.
\bjournal{Animal Behavior}
\bvolume{72}
\bpages{353--363}.
\bptok{imsref}%
\end{barticle}
%
\endbibitem

\bibitem[\protect\citeauthoryear{Wickelmaier and Schmid}{2004}]{Wickelmaier04}
%
\begin{barticle}[auto:STB|2012/06/04|06:16:18]
\bauthor{\bsnm{Wickelmaier},~\bfnm{F.}\binits{F.}} \AND
\bauthor{\bsnm{Schmid},~\bfnm{C.}\binits{C.}}
(\byear{2004}).
\btitle{A Matlab function to estimate choice model parameters from
paired-comparison data}.
\bjournal{Behavior Research Methods, Instruments, and Computers}
\bvolume{36}
\bpages{29--40}.
\bptok{imsref}%
\end{barticle}
%
\endbibitem

\bibitem[\protect\citeauthoryear{Yan, Yang and Xu}{2012}]{Yan11}
%
\begin{bmisc}[auto:STB|2012/06/04|06:16:18]
\bauthor{\bsnm{Yan},~\bfnm{T.}\binits{T.}},
\bauthor{\bsnm{Yang},~\bfnm{Y.}\binits{Y.}} \AND
\bauthor{\bsnm{Xu},~\bfnm{J.}\binits{J.}}
(\byear{2012}).
\bhowpublished{Sparse paired comparisons in the Bradley--Terry model.
\textit{Statist. Sinica} \textbf{22} 1305--1318}.
\bptok{imsref}%
\end{bmisc}
%
\endbibitem

\bibitem[\protect\citeauthoryear{Zermelo}{1929}]{Zermelo29}
%
\begin{barticle}[mr]
\bauthor{\bsnm{Zermelo},~\bfnm{E.}\binits{E.}}
(\byear{1929}).
\btitle{Die {B}erechnung der {T}urnier-{E}rgebnisse als ein {M}aximumproblem
der {W}ahrscheinlichkeitsrechnung}.
\bjournal{Math. Z.}
\bvolume{29}
\bpages{436--460}.
\bid{doi={10.1007/BF01180541}, issn={0025-5874}, mr={1545015}}
\bptok{imsref}%
\end{barticle}
%
\endbibitem

\bibitem[\protect\citeauthoryear{Zhao and Joe}{2005}]{Zhao05}
%
\begin{barticle}[mr]
\bauthor{\bsnm{Zhao},~\bfnm{Yinshan}\binits{Y.}} \AND
\bauthor{\bsnm{Joe},~\bfnm{Harry}\binits{H.}}
(\byear{2005}).
\btitle{Composite likelihood estimation in multivariate data analysis}.
\bjournal{Canad. J. Statist.}
\bvolume{33}
\bpages{335--356}.
\bid{doi={10.1002/cjs.5540330303}, issn={0319-5724}, mr={2193979}}
\bptok{imsref}%
\end{barticle}
%
\endbibitem

\end{thebibliography}
\end{document}